# Machine Learning on Neutron and X-Ray Scattering

Zhantao Chen[1,2†], Nina Andrejevic[1,3†], Nathan Drucker[1,4], Thanh Nguyen[1,5], R Patrick Xian[6], Tess Smidt[7], Yao Wang[8], Ralph Ernstorfer[6], Alan Tennant[9], Maria Chan[10], and Mingda Li[1,5*]

[1]Quantum Matter Group, MIT, Cambridge, MA 02139, USA

[2]Department of Mechanical Engineering, MIT, Cambridge, MA 02139, USA

[3]Department of Materials Science and Engineering, MIT, Cambridge, MA 02139, USA

[4]Department of Applied Physics, School of Engineering and Applied Sciences, Harvard University, Cambridge, MA 02138, USA

[5]Department of Nuclear Science and Engineering, MIT, Cambridge, MA 02139, USA

[6]Fritz Haber Institute of the Max Planck Society, 14195 Berlin, Germany

[7]Computational Research Division and Center for Advanced Mathematics for Energy Research Application, Lawrence Berkeley National Laboratory, Berkeley, CA 94720, USA

[8]Department of Physics and Astronomy, Clemson University, Clemson, SC 29634, USA

[9]Neutron Scattering Division, Oak Ridge National Laboratory, Oak Ridge, TN 37831, USA

[10]Center for Nanoscale Materials, Argonne National Laboratory, Lemont, IL 60439, USA

## Abstract

Neutron and X-ray scattering represent two state-of-the-art materials characterization techniques that measure materials' structural and dynamical properties with high precision. These techniques play critical roles in understanding a wide variety of materials systems, from catalysis to polymers, nanomaterials to macromolecules, and energy materials to quantum materials. In recent years, neutron and X-ray scattering have received a significant boost due to the development and increased application of machine learning to materials problems. This article reviews the recent progress in applying machine learning techniques to augment various neutron and X-ray scattering techniques. We highlight the integration of machine learning methods into the typical workflow of scattering experiments. We focus on scattering problems that faced challenge with traditional methods but addressable using machine learning, such as leveraging the knowledge of simple materials to model more complicated systems, learning with limited data or incomplete labels, identifying meaningful spectra and materials' representations for learning tasks, mitigating spectral noise, and many others. We present an outlook on a few emerging roles machine learning may play in broad types of scattering and spectroscopic problems in the foreseeable future.

[†]These authors contribute equally to this work.

*Corresponding author: mingda@mit.edu.

## Contents

# I. Introduction

## I.1. Neutron and X-ray scattering in the data era

Neutron and X-ray scattering are two closely related and complementary techniques that can be used to measure a wide variety of materials' structural and dynamical properties, from atomic to mesoscopic scales[1,2]. Representing two state-of-the-art materials characterization techniques, neutron and X-ray scattering have witnessed significant advancement in the past several decades. As the average neutron flux reached a plateau $\sim 10^{15}\,\mathrm{n/cm^2/s}$ for reactor-based neutron generation, accelerator-based neutron generation has improved steadily (***Figure 1*****a**)[3]. The planned Second Target Station (STS) at Oak Ridge National Lab (ORNL) has a 25x enhancement in brightness and a factor of 10-1000 capability enhancement in instruments comparing other neutron sources in the US. For X-ray scattering, the peak brightness of synchrotron sources has increased drastically across a broad range of X-ray photon energies (***Figure 1*****b**)[4]. In fact, the improvement in peak brightness of synchrotron X-ray sources even exceeds the rate of Moore's law (***Figure 1*****c**)[5,6], with a few major facility upgrades such as APS-U, ESRF-EBS, and PETRA-IV, bringing significant capability boosts. A direct consequence of the enhanced capability is the high efficiency of data collection, enabling the measurement of more diverse types of materials.

In addition to increased data availability for a broader materials composition space, the higher brightness further opens up the possibility for higher-dimensional data collection for a single material type or within one scattering experiment. Spectroscopies like time-of-flight inelastic neutron scattering measure the dynamical structure factor in four-dimensional (4D) momentum-energy $(\mathbf{Q},\omega)$ space, while X-ray photon correlation spectroscopy measures the intensity auto-correlation in 4D momentum-time $(\mathbf{Q},t)$ space[7]. The emerging frontier of multimodal scattering, which simultaneously measures samples with multiple probes, or in *in-situ* environments such as extreme temperature or pressure, elastic strain, or applied electrical and magnetic fields, introduces additional dimensions to the measured parameter space. Alongside high intrinsic momentum $\mathbf{Q}$, energy $\omega$, and time $t$ dimensions, multimodality leads to an even higher overall data dimension and adds inevitable complexities to data analysis.

Lastly, the discovery of new functional and quantum materials – often accompanied by novel or unexpected emergent properties – poses a significant challenge to materials analysis. In many scattering experiments with a given measurable signal $S_{\exp}(\mathbf{Q}, w, t, \ldots)$, there exist associated theoretical models $S_{\text{model}}(\mathbf{Q}, \omega, t, \ldots; \boldsymbol{\theta})$, parameterized by a set of fitting parameters $\{\boldsymbol{\theta}\}$ representing materials properties to extract. For the optimal fitting parameters $\boldsymbol{\theta} = \boldsymbol{\theta}_{op}$, the difference between the experiment $S_{\exp}$ and model $S_{\text{model}}$ reaches a minimum, i.e.

$$\boldsymbol{\theta}_{op} = \arg\min_{\boldsymbol{\theta}} \left\| S_{\exp}(\mathbf{Q}, \omega, t, \ldots) - S_{\text{model}}(\mathbf{Q}, \omega, t, \ldots; \boldsymbol{\theta}) \right\|.$$

However, even for a perfect fitting $S_{\exp}(\mathbf{Q}, \omega, t, \ldots) \equiv S_{\text{model}}(\mathbf{Q}, \omega, t, \ldots; \boldsymbol{\theta}_{op})$, the information that can be extracted is still ultimately limited by the theoretical model itself. Until recently, avenues that can access materials' properties outside the parameter set $\{\boldsymbol{\theta}\}$ have been lacking.

In short, large data volume, high data dimension combined with multimodality, and new classes of quantum and functional materials with emergent properties that go beyond approximate models, all call for a revolutionary approach to learn materials properties from neutron and X-ray scattering data. Machine learning[8,9], especially emerging techniques that incorporate physical insights[10-15] or respect symmetries and physical constraints of atomic, crystalline, and molecular structures[16-26], appears to be a promising and powerful tool to extract useful information from large, high-dimensional datasets, going far beyond approximate models. The past few years have witnessed a surge in machine learning research with scattering and spectroscopic applications. Even so, we foresee that machine learning, if properly implemented, has the potential to not only serve as a powerful tool to do data analysis, but also to gain new knowledge and physical insights of materials, which can assist experimental design and accelerate materials discovery.

### I.2. Integrating machine learning into the scattering setup

Machine learning has already been widely applied to materials science in many aspects, especially in directly predicting or facilitating predictions of various materials properties from structural information, including but not limited to mechanical properties[24,26-28], thermodynamic properties[27,29-31], and electronic properties[24,32-38]. The strong predictive power and representation learning ability of machine learning models can lead to much lower computational cost compared to expensive numerical methods like first-principles calculations but with comparable accuracy. This feature greatly accelerates the materials discovery and design[39-44]. Machine learning models can also be trained to learn interatomic force fields and potential energy surfaces[45-50], where the accurate yet computationally-cheap access to atomic potentials has proven successful in simulating the transitions in a disordered silicon system with 100,000 atoms[51]. Machine learning models have already initiated a paradigm shift in the way people study materials science and physics[52-58].

To see how machine learning can be applied to neutron and X-ray scattering, we show a simple scattering setup in ***Figure 2*****a**. A beam of neutrons or X-ray photons is generated at the source. After passing through the beam optics that prepares the incident beam state, the beam impinges on the sample with a set of incident parameters $(I_i, \mathbf{k}_i, E_i, \varepsilon_i, ...)$, where $I_i, \mathbf{k}_i, E_i$, and $\varepsilon_i$ denote the incident beam intensity, momentum, energy, and polarization, respectively. After interacting with the sample, the scattered beam can be described by another set of parameters $(I_s, \mathbf{k}_s, E_s, \varepsilon_s, ...)$, which are partially or fully recorded by the detector. In this source-sample-detector tripartite scheme, the possible application scope of machine learning can be seen clearly: At the "source" stage, machine learning can be used to optimize beam optics; at the "sample" stage, machine learning can be used to better learn materials properties; while at the "detector" stage, machine learning can be used to improve data quality, such as realizing super-resolution. Setting aside the "source" and "detector" stages, which will be introduced in Section IV, we focus on the "sample" stage, particularly the application of machine learning to relate materials' spectra and their properties.

To further illustrate the general relationship between machine learning and scattering spectra, we consider the scattering data as one component in a typical machine learning architecture. In the case of supervised machine learning, the scattering spectral data can serve either as input to predict other materials properties (***Figure 2*****b**), or as output generated from known or accessible materials parameters, such as atomic structures and other materials representations (***Figure 2*****c**). Alternately, unsupervised machine learning can be used to identify underlying patterns in spectral data through dimensionality reduction and clustering, which can be useful for data exploration or identification of key descriptors in the data (***Figure 2*****d**).

## I.3. Machine learning architectures for scattering data

With the various roles machine learning may play in a scattering experiment pipeline, one may ask what particular machine learning architecture should be used for a certain task. Given the no free lunch theorem for optimization[59], many algorithms are interchangeable. Even so, a number of machine learning models are naturally suited to scattering experiments. Here we introduce a few categories of useful architectures, many of which are implemented in the examples that will be discussed in later sections.

*Representation of materials.* For materials studies, the representation of materials, particularly atomistic structures, is crucial. Various representational approaches have been developed to describe molecules and solids. These methods include the Coulomb matrix representation[60], which translates molecules into matrices, the Ewald sum matrix representation, which generalizes Coulomb matrix to infinitely periodic structures[61], partial radial distribution function (PRDF), which describes radial density of some species around an atom[16], the atom-centered symmetry functions which contain both radial and angular information[62], and the smooth overlap of atomic positions (SOAP) and power spectrum[63]. A review on materials representation can be found in Ref [54].

*Representation of scattering data*. Paired with materials representation is the representation of scattering data. The scattering intensity can be stored as a high-dimensional array $I \in \mathbb{R}^{N_{\mathbf{k}} \times N_{\omega} \times N_{\varepsilon}}$ indexed by momentum **k**, energy $\omega$, and polarization $\varepsilon$. Such data structures are naturally compatible with convolutional neural networks (CNN), which has been widely applied in image processing. Moreover, atomic structures can also be interpreted as images by computing density fields in 3D real-space grids based on atomic species and positions, which enables them to work with convolutional filters[43,64]. Architectures beyond CNN, such as deep U-Net, also exist, which decreases the feature size while increasing the feature numbers, then performs the inverse operation with skip connections enabled between corresponding levels (**Figure 3a**)[65].

*Autoencoder and generator.* Another useful architecture is the variational autoencoder (VAE)[66], which compresses the input into some distributed area in a lower-dimensional latent space (encoding), followed by the optimized recovery of input from the low-dimensional representation (decoding). The latent space is thus a "compressed" continuous representation of the training samples, which can be very useful in learning representations for materials (**Figure 3b**). For example, VAE can be combined with CNN to learn latent representations of atomic structures[64]. Moreover, the stability of crystal structures can be easily inferred from latent space clustering[43], and similar ideas can also be applied to analyze scattering data, such as X-ray absorption spectroscopy[38] or neutron diffuse scattering[67]. Another use of VAE is to serve as generative models to facilitate material design, such as generating new structures through sampling and exploring the latent space[43]. The generative adversarial network (GAN) is another popular generative framework that is composed of a generator and a discriminator (**Figure 3c**)[68]. The generator is a neural network that converts latent space representations to desired objects such as crystal structures[44], while the discriminator is another network that aims to discern "fake" (generated) from "realistic" (training) samples. The main goal of the generator is to create high-fidelity objects that can pass through the discriminator test.

*Graphic neural networks.* Graph neural networks with nodes and edges are naturally suited to represent atomic structures, where atoms can be represented as nodes, and their bonds correspond to edges in a graph. In graphs, information at each node is updated with filtered information from its neighboring nodes, mimicking the local chemical environment where an atom is most influenced by neighboring atoms. The crystal graph CNN (**Figure 3d**)[24] and the Euclidean neural network ($E^3NN$)[19,69,70], are two such examples. $E^3NN$ is equipped with sophisticated filters that incorporate radial functions and spherical harmonics, where equivariances in 3D Euclidean space are guaranteed, and all crystallographic symmetries of input structures are preserved. Consequentially, $E^3NN$ can augment data from a symmetry perspective without increasing the data volume.

*Non-parametric learning algorithms.* The aforementioned machine learning architectures contain parameters that need to be learned during the training process, yet there exist plenty of unsupervised learning or non-parametric algorithms which do not contain learnable parameters but are more procedural. For instance, *k*-means clustering and Gaussian mixture models (GMMs) can be applied to data clustering, decision trees such as gradient boost trees (GB Trees) can be applied for classification and regression (**Figure 3f**), and principal component analysis (PCA) can be used for data dimension reduction. One particularly interesting method is the non-negative matrix factorization (NMF), which decomposes a matrix into lower dimensions but maintains an intuitive representation composed of different parts (**Figure 3e**)[71,72]. Conceptually, NMF resembles the widely used dimension reduction algorithm of PCA, but with additional non-negativity constraints. Such non-negative matrix descriptions are extremely powerful when interpreting some physical signals like music spectrograms[73]. Likewise, scattering data collected from detectors have non-negative counts in the array $I \in \mathbb{R}^{N_{\mathbf{k}} \times N_{\omega} \times N_{\varepsilon}}$ and can be decomposed with NMF in principle.

## II. Static Properties in Reciprocal or Real Space

### II.1. Diffraction with machine learning

To see how machine learning can benefit neutron and X-ray diffraction, we follow the taxonomy in ***Figure 2***: in supervised learning, diffraction can serve as an input to predict materials properties, or predict the structure itself, while in unsupervised learning, diffraction can be used to perform classification without additional data labels. The inverse problems, such as using structures or other physical properties to predict diffraction patterns, either belong to physics-based forward problems or have less value for machine learning studies and will be left out of this discussion.

*Diffraction or structure as input, property as output*. Since the most straightforward information extractable from diffraction is atomic structure, whose variation is directly associated with mechanical properties, we start by discussing an example that shows how diffraction can be used to predict elastic constants in complicated materials, taking high-entropy alloys as an example. High-entropy alloys have received tremendous attention in the past decade due to their extraordinary strength-to-weight ratios and chemical stability. However, given their complex atomic configurations, direct property calculation has been challenging. To enable efficient prediction of elastic constants in high-entropy alloys, Kim *et al.* conduct a combined neutron diffraction, *ab initio* calculation, and machine learning study[74]. In this study, an *in-situ* diffraction experiment and high-quality *ab initio* density functional theory (DFT) calculations with special quasi-random structure (SQS) are carried out on the high-entropy alloy $Al_{0.3}CoCrFeNi$. The experimental result and *ab initio* calculations of elastic constants show good agreement and thus can serve as the ground truth (***Figure 4*****a**, “ground truth” block). Due to the limited neutron beamtime and high computational cost of DFT+SQS, it would be unrealistic to either measure or compute the elastic constants in a large number of high-entropy alloys. To bridge this key gap, the authors build a GB Tree-based predictive model using a separate set of nearly 7,000 ordered, crystalline solids from the Materials Project, in which the elastic constants have already been properly labeled (***Figure 4*a**, "machine learning" block). It is worth mentioning that the training set and validation set do not contain any high-entropy alloys. Even so, there are a few indicators that demonstrate the model's transferability and generalizability. On the one hand, the elastic constants of $Al_{0.3}CoCrFeNi$ predicted by the machine learning model show good agreement with "ground-truth" values established by experiments and DFT+SQS calculations. On the other hand, the lower training error compared to a benchmark model, and the reasonable dependence as a function of training data volume give a level of confidence in the model generalizability.

The high-entropy alloy example demonstrates a general pathway for efficient property predictions in complex materials, where data scarcity is a common challenge. With a small labeled "hard" dataset $\{x_i^H, y_i^H\}$, $i = 1,2,...,n$ that is hard to acquire, a direct machine learning may not be feasible due to the low data volume *n*. To make machine learning possible, another large set of $\{x_j^E, y_j^E\}$, with $j = 1,2,...,N$ and $N >> n$, can be used, where the features $x_j^E$ and labels $y_j^E$ form another set that is easier to obtain, such as from simple crystalline solids or from efficient forward computation. The key step is to build a predictive model using the large, "easy" set $\{x_j^E, y_j^E\}$ that also minimizes the test error from $\{x_i^H, y_i^H\}$ (***Figure 4*b**). A few approaches can help to achieve this step, such as transfer learning and data augmentation. The outcome thus has a level of generalizability. In the previous example,

$$
\begin{aligned}
x^H &= diffraction / structure\ in\ high\ entropy\ alloys, \\
y^H &= elastic\ constants\ in\ high\ entropy\ alloys, \\
x^E &= structure\ in\ crystalline\ solids, \\
y^E &= elastic\ constants\ in\ crystalline\ solids,
\end{aligned}
$$

and $\{x_i^H, y_i^H\}$ is simply used to test the machine-learning model built upon $\{x_j^E, y_j^E\}$ (***Figure 4*b,** "right" arrow).

*Diffraction as input, structure as output.* In addition to the direct structure-to-property prediction, given the close relationship between diffraction and structure, another type of machine learning problem is to perform diffraction-to-structure prediction, as done by Garcia-Cardona *et al.* for neutron diffraction[75]. The conventional solution to this problem is iterative optimization of the computed scattering patterns from physics-based forward models, which leaves room for machine learning training. A key challenge, however, lies in the scarcity of labeled neutron diffraction data, that

$$x^H = experimental\ neutron\ diffraction,$$
$$y^H = y^E = atomic\ structure.$$

To facilitate machine learning training, simulated diffraction patterns are generated by sweeping the structure parameter space (lattice parameters, unit cell angles etc.), i.e., $x^E =$ *simulated neutron diffraction*. This example takes advantage of data augmentation (***Figure 4*b,** "left" arrow). Similarly, augmented X-ray diffraction (XRD) data can be used to obtain crystal dimensionality and space group information[76]. Another example uses the atomic pair distribution function (PDF) to predict space group information[77]. Conventionally, PDF is powerful in determining the local order and disorders information[78]. By setting

$$x^E = simulated\ PDF,$$
$$y^E = y^H = space\ group,$$

PDF is empowered by allowing the determination of the global space group information. A separate set $x_i^H =$ experiment PDF containing 15 examples are used as a test (***Figure 4*b,** "right" arrow), where 12 of them have their space group appearing in the top-6 predicted labels.

*Unsupervised learning of diffraction.* Besides supervised learning on XRD or PDF spectra, another boost for scattering data lies in unsupervised learning, which seeks the internal categorical data structure. Since conventional fitting and refinement methods have been maturely developed to identify phases among different crystallographic structures, one key application for unsupervised learning is phase identification in a complex compositional phase space where multi-phases coexist. One key milestone in the unsupervised learning algorithm for XRD analysis is the NMF, which can decompose the spectra into simpler basis patterns. Recalling the introduction in section I.3, NMF decomposes a non-negative matrix $Y \in \mathbb{R}^{M \times N}$ into two smaller non-negative matrices, namely the basis matrix $A \in \mathbb{R}^{M \times K}$ representing $K$ basis patterns and the coefficient matrix $X \in \mathbb{R}^{K \times N}$ indicating contributions of those patterns. In the XRD context, $Y_{mn}$ denotes the diffraction intensity of *m*-th composition at the *n*-th sample diffraction angle (or equivalently momentum transfer *Q*). Long *et al.* apply NMF to identify phases within the metallic Fe-Ga-Pd ternary compositional phase diagram[79]. For instance, given a nominal composition $Fe_{46}Pd_{26}Ga_{28}$, NMF decomposes its XRD pattern (***Figure 5*****a**, middle figure) into a weighted sum of 5 basis patterns (***Figure 5*****a**, bottom and top figures) with a total number of $K_{\mathrm{NMF}} = 9$ basis patterns. The entire structural phase diagram in the compositional phase space can thereby be constructed (***Figure 5*****b**), which contains the quantitative weight information of each pure constituent phase. Limitation exits, though, when the same nominal composition corresponds to different structure combinations with slightly varied diffraction peaks. To overcome this limitation, Stanev *et al*. extend NMF with custom clustering (NMFk) algorithms to capture the nuance peak shifts from lattice constant change, which can further resolve the constituent phases even within the same nominal composition[80]. As compared to NMF, where the small matrix dimension *K* is chosen manually with trial-and-error, NMFk automatically searches and optimizes *K*. The same Fe-Ga-Pd dataset first analyzed in Ref [79] by NMF is re-analyzed using NMFk[80]. An optimized basis pattern number $K_{\mathrm{NMF}} = 13$ is found by NMFk, which contains 4 basis patterns representing BCC Fe structures but with a slight peak shift (***Figure 5*****d**). Although the BCC Fe structure corresponds to almost identical regions in the structural phase diagrams produced by NMFk and NMF methods (Comparing ***Figure 5*****c** to ***Figure 5*****b**, blue color), the weight of each NMFK basis pattern of BCC Fe structures can be seen clearly (***Figure 5*****e**), tracing the nuanced lattice parameter change in the phase diagram. Besides being applied to alloy phase diagrams, in a more recent example, XRD measurement is analyzed to obtain the quantum phase diagram with charge ordering and structural phase transitions, with a novel approach called XRD temperature clustering (X-TEC) building upon the GMM[81].

Beyond learning materials properties and seeking for structure-property relations, machine learning has also been applied to empower the analysis of the diffraction patterns themselves[82-86]. Since the focus here is to explore materials properties, we leave those examples to Section IV.2 as part of the section on the data analysis process.

## II.2. Small-angle neutron and X-ray scattering

Small-angle scattering (SAS), including small-angle neutron and X-ray scattering (SANS and SAXS) are very powerful techniques to probe structures and their evolution on the scale of 0.5 nm to 100 nm[87,88], and have been widely applied to study soft matter systems like rough surfaces[89], colloids and polymers[90-92], biological macromolecules[92-95], and mesoscopic magnetic structures such as magnetic vortex lattices in superconductors (SANS only)[96-99]. In the past few years, a surge of machine-learning augmented SAS works have been reported[100-112]. At least two reasons make SAS an ideal technique to benefit from machine learning. On the one hand, SAS represents one of the rare kind of techniques where experimental data can directly and quantitatively compare with theoretical models, with minimal experimental data post-processing needed. This direct data-to-data comparison increases transferability using computational "easy" data $\{x_j^E, y_j^E\}$ to do training, with high quality. On the other hand, SAS allows for highly efficient synthetic data generation, since in many cases, only effective geometrical models at intermediate scales are needed to compute 1D SAS spectra of $I(Q)$. Even in the cases of atomistic-scale data generation, methods with low computational cost, such as molecular dynamics, Monte Carlo simulations, or micromagnetic simulations, are generally sufficient without carrying out the full *ab initio* calculations.

*Spectra as input, structure as output.* Since the original goal for SAS is to learn structural information, we start by introducing one example that predicts structural properties. Franke *et al.* provide such a machine-learning-based structure predictor in bio-macromolecular solutions[102]. For a given geometrical object, although the form factor and the corresponding SAS spectra are directly computable (***Figure 6*a**), the effect of disorder must be considered to generate data that are close to reality. To consider the disorder effect, an ensemble optimization method is implemented to generate SAS patterns with random chains followed by averaging to simulate mixtures (***Figure 6*b**), which augment the original geometrical data (***Figure 6*e**, orange block). The first task is to classify the shape of macromolecules from SAS. Defining a structural parameter called radius of gyration $R_g$, the original SAS can be compressed into a 3D parameter space with 3 coordinates representing normalized apparent values $V'$ corresponds to the integral upper bounds of $|\mathbf{Q}|R_g = 3, 4, 5$, respectively. It can be seen directly that different basic shapes separate well in this 3D parameter space (***Figure 6*****b**). By performing unsupervised *k*-nearest neighbor classification, the shapes with mixture and disorders can also be classified from SAS curves. To perform structural parameter prediction, a separate set of atomistic structure data from protein database (PDB) is used to compute both SAS patterns and structural parameters, from which a predictive machine-learning model is built, showing good transferability when applied to experimental database (***Figure 6*****d**). A summarized workflow is shown in ***Figure 6*****e**. In this example, we can still state

$$x^H = experimental\ SAS,$$
$$x^E = simulated\ SAS,$$
$$y^H = y^E = shape\ and\ structural\ characteristics,$$

but with two different sources

$$\begin{matrix} x^{E1} = geometrically\ simulated\ SAS \\ x^{E2} = atomistically\ simulated\ SAS \end{matrix},$$

highlighting the different focus in obtaining shape features and structural parameters, respectively. The goal of shape classification and structure parameter prediction, with different synthetic data augmentation for respective tasks, represents an active area using machine learning on SAS[106,111], which has been applied systems like RNA[108] and 3D protein structures[110]. Machine learning also enables a direct analysis using 2D SAS data[105,112], where traditional analysis frequently needs a data reduction to 1D for further analysis.

Another example of machine learning applied to SAS is in micromagnetic structural determination from SANS. As in the soft matter cases, real space structural information is encoded in 2D maps of the neutron scattering cross-section. As noted previously, a strong benefit of magnetic SANS is that the structure factor and cross-section are relatively straightforward to calculate from a theoretical model – often a micromagnetic continuum model – of the real space magnetization. With sufficiently labeled experimental data and micromagnetic simulations, supervised learning with neural networks or unsupervised clustering methods can be used to solve the inverse scattering problem of determining real space magnetic structures from SANS spectra at mesoscale.

*Spectra as input, other property as output.* Since the structures of macromolecules are directly linked to their microscopic interactions, one further use of machine learning is to augment SAS to learn interatomic interaction properties. Demerdash *et al.* directly extract force-field parameters using SAS[103], which is a feedback loop "force-field parameters → molecular dynamics (MD) simulations → SAS calculations → force-field parameters" (***Figure 6*f**). Such refined force-field parameters improve the agreement between simulated SAS and certain experimental data.

## II.3. Imaging and tomography

Neutron[113,114] and X-ray[115] imaging encompass a variety of modalities and have become essential techniques to unravel multidimensional and multiscale information in materials systems. As the complexity and size of imaging data grows, machine learning has also been applied to solve a variety of imaging-related computational tasks, including tomography and phase-contrast imaging. Tomography and phase-contrast imaging are two types of high-dimensional imaging concerning the beam absorption and phase-shift associated with sample rotation, respectively. We restrict the further discussion to materials science and refer the readers to other reviews for applications in biomedical imaging[116,117]. Despite the variety of imaging modalities, the major data processing steps generally include image reconstruction and image segmentation.

In image reconstruction, one recovers the real space information (usually the amplitude and phase of the imaged object) from data obtained at different sample positions. Neural network-based reconstruction algorithms have been shown to improve the reconstruction speed and quality[118], as demonstrated in a neutron tomography experiment[119]. Yang *et al.* demonstrate the use of GAN for reconstructing limited-angle tomography data by casting the reconstruction into an image translation problem between the sinogram domain and the real space[120]. Their method, called GANrec, has been shown to tolerate large missing wedges without obvious reconstruction quality degradation. GANrec has been successfully applied to tomographic imaging of zeolite particles deposited on a microelectromechanical systems (MEMS) chip, which, due to limited rotation capability, has a missing wedge of 110°. The reconstruction from GANrec shows significant improvement over outcomes from conventional reconstruction algorithms, which are corrupted by artifacts due to the missing data.

As to image segmentation, that to separate pixels representing the desirable structures from the background, the classical architectures include variants of CNN and U-Net[65]. A number of related studies have been conducted, such as materials' defect recognition[121-123], mineral phase segmentation[124], automated feature extraction for micro-tomography[125], and non-destructive, *in vivo* estimation of plant starch[126]. Deep transfer learning, which has demonstrated great power in image processing, can also be applied to feature extraction in X-ray tomography[127], where a pre-trained network on large image database plays similar role to $\{x_j^E, y_j^E\}$ following the idea in ***Figure 4*****b**. Deep U-Net, on the other hand, can be used to perform image segmentation beyond CNN[128].

A particularly powerful technique, called the coherent diffraction imaging (CDI), has attracted significant research attention since its first demonstration in 1999[129]. Contrary to conventional imaging, the resolution in CDI is not limited by the imaging optics. This allows the 3D structure determination in nano-sized materials through computational phase retrieval[130,131]. Given the complex data relations, absent phase information, and high data volume, machine learning is becoming a promising tool for CDI analysis. As an example, the shapes of helium nano-droplets have been measured by single-shot CDI from free-electron laser[132], where the shape classification from diffraction images can be realized through a CNN[133]. More recently, Cherukara *et al.* construct a CNN with two output branches (***Figure 8*****a**) that aims to directly solve the phase retrieval problem in a particular type of CDI called ptychography, where the sample is scanned through[134]. By inputting the diffraction patterns at different spatial scan points (row A in ***Figure 8*****b**), the retrieved amplitude and phase images of the 2D tungsten calibration chart sample obtained from machine learning (row C and E in ***Figure 8*****b**) show good agreement with the conventional iterative phase retrieval algorithm approach (row B and D in ***Figure 8*****b**). The CNN-assisted approach can speed up the scanning effectively by 5 times, thus greatly reducing the imaging time and lowering the dose. On the other hand, Scheinker and Pokharel build an additional model-independent adaptive feedback loop on top of the CNN output[135], which allows for more accurate recovery of the 3D shape (***Figure 8*****c**). Iterative projection approaches still demonstrate great flexibility in tomographic reconstruction because constraints such as multiple scattering effects can be well captured from physical models[136], while currently only implemented for limited cases via machine learning-based approaches in optical imaging[137].

# III. Spectroscopies and Dynamical Properties

## III.1. X-ray absorption spectroscopy

X-ray absorption spectroscopy (XAS) is another characterization technique widely-used in materials science, chemistry, biology and physics. The possibility to reach excellent agreement between experimental and computational data makes XAS suitable for training machine learning models on bulk computational data that translate well to experimental examples. The absorption of X-rays reflects electronic transitions from an atomic core orbital to either unoccupied bound levels or the free continuum, producing sharp jumps in the absorption spectrum at specific energies called absorption edges[138]. Such a measurement is therefore sensitive to the species of the absorbing atom, as well as to its valence state and local chemical environment, including the local symmetry, coordination number, and bond length[139,140]. As a result, XAS is routinely used in the characterization of materials' structural and electronic properties. However, interpretation of XAS spectra ranges from qualitative comparisons with known model complexes, to more quantitative comparisons with theoretical models[141,142] or bandstructure calculations, making the process difficult to standardize and automate across materials and applications. Machine learning methods are therefore sought to better extract and decipher the rich electronic and structural information encoded in XAS signatures.

To address this key objective, Carbone *et al*. develop a neural network classifier to identify the local coordination environments of absorbing atoms in more than 18,000 transition metal oxides using simulated K-edge X-ray absorption near-edge structure (XANES) spectra[143]. The input of their neural network model is the discretized XANES spectrum, while the output is a predicted class label corresponding to one of three coordination geometries: tetrahedral, square pyramidal, and octahedral. The authors achieve an average 86% classification accuracy when using the full (pre-, main-, and post-edge) feature space of the discretized spectra; however, by also training their model using only the pre-edge region, they further reveal the significance of features beyond the pre-edge for accurate classification of the coordination environments (**Figure 7a**). The work of Torrisi *et al*. expands upon this approach by subdividing the discretized XANES spectra into smaller domains ranging from 2.5 eV to 12.5 eV, thereby capturing spectral features on both coarse and fine scales[144]. The spectrum within each domain is then fit by a cubic polynomial whose coefficients serve as inputs to random forest models for predicting the properties of interest, including coordination number, mean nearest-neighbor distance, and Bader charge. Through this multiscale featurization, the authors highlight the importance of developing effective data representations to improve model interpretability and accuracy.

The role of data representation is also explored by Madkhali *et al.* for the inverse problem; that is, how the choice of representation for the local environment of an absorbing atom affects the performance of a neural network in predicting the corresponding K-edge XANES spectrum[145]. In particular, the authors examine two different representations of chemical space, the Coulomb matrix and radial distribution curve (RDC), shown in ***Figure 7*b**), to represent the local environment around an Fe absorption site, and evaluate them based on ability to recover the Fe K-edge XANES spectra of 9040 unique Fe-containing compounds. They conclude that RDC featurization can achieve smaller mean squared error (MSE) between the predicted and target XANES spectra more quickly and with fewer data samples, reinforcing the need for effective data representations of materials-specific descriptors.

Another focus of machine learning efforts in this context includes accelerating high-throughput modeling of XAS spectra. As a proof of concept, Carbone *et al*. show that a message-passing neural network (MPNN) is capable of predicting the discretized XANES spectra of molecules to quantitative accuracy by using a graph representation of molecular geometries and their chemical properties[146]. An MPNN, shown in **Figure *7*c**, refers to a neural network framework that operates on graph-structured data: Hidden state vectors at each node in the graph are updated according to a function of their neighbors' state vectors for a specified number of time steps, and the results are ultimately aggregated over the entire graph to produce the final output[147]. The structural similarities between MPNNs and molecular systems suggest that these networks may better predict molecular properties by remaining invariant to the molecular symmetries that help determine these properties. In their work, Carbone *et al*. construct each molecular graph by associating with each graph node a list of atom features (absorber, atom type, donor or acceptor states, and hybridization) and with each graph edge a list of bond features (bond type and length). The MPNN then passes the encoded feature information between adjoining nodes to learn effective atomic properties before computing a discretized output XANES spectrum from the final hidden state vectors. The network is optimized by minimizing the mean absolute error between this predicted spectrum and a ground-truth XANES spectrum obtained from simulation. By contrast, Rankine *et al*. implement a deep neural network to estimate Fe K-edge XANES spectra, relying only on geometric information about the Fe local environment as input[148]. Specifically, the authors represent the local environment around the Fe absorption site by computing a discrete RDC comprising all two-body pairs within a fixed cutoff radius. Despite the limited input information, they demonstrate that a properly trained network can be used to make rapid, quantitatively accurate predictions while circumventing the time and resource demands of advanced theoretical calculations.

Lastly, one major advantage of XAS is its compatibility with diverse samples, both crystalline and amorphous, and sample environments, as in the case of *in situ* or *operando* measurements under extreme temperatures or externally applied fields, leading to diverse applications and opportunities for machine learning-assisted analysis. In particular, XAS is a prominent method used to correlate the structure of nanoparticle catalysts to properties such as catalytic activity, which is often characterized under the *operando* conditions of a harsh reaction environment, shown in **Figure 7d**. Thus, the predictive ability of machine learning methods is attractive for directly recognizing encoded structural descriptors, such as coordination number, from evolving XAS spectral features. For example, Timoshenko *et al*. demonstrate that neural networks can be used to predict the average coordination numbers of Pt nanoparticles directly from their XANES spectra, which can then be used to determine particle sizes, shapes, and other structural motifs needed to inform catalyst design[149]. Several successful examples of machine learning-aided analysis for *operando* XAS spectra of catalyst structures have been reported in recent years[150-153]. Machine learning has also been applied to conduct high-throughput screening and obtain additional chemical insight into the atomic configurations of thin films monitored by *in situ* XAS during synthesis[154]. Overall, machine learning methods have shown incredible potential for improving and accelerating the analysis of this versatile characterization tool, and more widespread integration of machine learning solutions within routine XAS analysis workflows may be on the horizon.

### III.2. Photoemission spectroscopies

Contrary to XAS, which is generally bulk sensitive, there is another surface-sensitive, photon-in, electron-out technique, named photoelectron or photoemission spectroscopy (PES), measured with light sources from hard X-ray to extreme ultraviolet (UV) energy range. PES provides the direct access to a material's electronic structure[155,156]. The high sensitivity of X-ray photoelectron spectroscopy (XPS) to the chemical environment makes it an essential tool for composition quantification. In this regard, machine learning-based spectrum fitting may be used to disentangle complex overlapping spectra. Aarva *et al.* use fingerprint spectra calculated with bonding motifs obtained from an unsupervised clustering algorithm to fit X-ray photoelectron spectra[157]. Drera *et al.* use simulated spectra to train a CNN to predict chemical composition directly from multicomponent X-ray photoelectron spectra from a survey spectra library[158]. Their approach obviates the need to fit these complex spectra directly, while showing robustness against the contaminant signal within the survey spectra.

Apart from chemical quantification, modern PES with momentum-resolving detectors is capable of mapping the entire electronic structure of materials through multidimensional detection of photoelectron energy and momentum distributions[155,159]. The resulting 4D intensity data in energy-momentum space from PES share the same data structure with the inelastic scattering for vibrational spectra. While this analogy implies transferability of machine learning approaches developed for inelastic scattering to be discussed in section III.3, the relation between PES observables and microscopic quantities is significantly more complex due to the quantum nature of the electronic states and the multiple prefactors that effectively modulate the intensity values in a momentum-dependent manner[160]. The current understanding of the complex photoemission spectra is limited by the available computational tools. Therefore, machine learning is a potential avenue to understand such data. Highlighting the dispersive features is of primary importance for comparison between experiments and theories. For this task, robust methods are needed to tolerate the noise level and intensity modulations in the data[161]. Peng *et al.* train a super-resolution neural network based on simulated angular resolved PES (ARPES) data, the "easy" set $\{x_j^E, y_j^E\}$, to enhance the dispersive features in experiment data, the "hard" set $\{x_j^H, y_j^H\}$, without explicit models of the band dispersion[162]. Xian *et al.* cast band fitting as an inference problem and use a probabilistic graphical model to recover the underlying dispersion[163]. Remarkably, this approach does not require training but a reasonably good prior guess as a starting point. Its reasonable computational scaling allows the reconstruction of multiband dispersions within the entire Brillouin zone, as demonstrated in 2D material tungsten diselenide ($WSe_2$).

### III.3. Inelastic scattering

One of the major triumphs of neutron and X-ray scattering is inelastic scattering, which measures the elementary excitations of materials[1,164-167]. There are generally two types of elementary excitations at milli-eV energy range, including a) collective atomic vibrations, such as phonons in crystalline solids[168-172] and Boson peaks in amorphous materials[173-177], and b) magnetic excitations, which are essential to understand the nature of strongly correlated materials[178], such as frustrated magnetism systems[179-181] and unconventional superconductors[182-184]. However, unlike elastic scattering, where massive synthetic data can be generated from forward models to build the "easy" set $\{x_j^E, y_j^E\}$, inelastic scattering is challenging for machine learning, due to the atomistic origin and quantum nature of the excitations, where forward models have high computational cost. Therefore, one major hurdle for machine learning inelastic scattering is data scarcity. Here we introduce two examples of using machine learning to overcome this hurdle to study elementary excitations of phonons and magnetic excitations, respectively.

For machine learning phonon studies, Chen, Andrejevic and Smidt *et al.* build a machine-learning-based model that predicts phonon density of states (DOS) by inputting atomic coordinates[185]. Two challenges exist in this problem. There has been a lack of large training set; a reliable density-functional perturbation theory (DFPT) database contains a small set of around 1,500 examples[186]. In addition, the predicted outcome, the phonon DOS, is a continuous curve instead of a scalar quantity. To tackle these challenges, a special graph neural network, termed Euclidean neural network[187], is implemented. The Euclidean neural network sees the equivariance from the index permutations, crystal rotations and translations, and thus fully respects the crystallographic symmetry (**Figure 9a**). The inherent symmetry effectively augments data without increasing its volume. Intuitively, the physical constraint imposed in the neural network acts like regularization, while encoding symmetry into the neural network resembles data augmentation of input data by symmetry operations. The predicted phonon DOS is shown in **Figure *9*b** with each of four rows representing an error quartile. For lower error predictions (first three rows in **Figure *9*b**), the fine shape of DOS can be well captured; for high-error predictions (fourth row in **Figure *9*b**), the coarse feature such as bandwidth and DOS gap can still largely be predicted. With such a predictive model available, the computational cost for phonon DOS is significantly reduced, and the prediction in alloy systems become feasible.

As to magnetic systems, Samarakoon *et al.* implement an autoencoder to assist the estimation of magnetic Hamiltonian parameters $\{J\}$ in spin ice $Dr_2Ti_2O_7$, including magnetic exchange coupling between neighborhood spins and magnetic dipolar interactions[67]. Although the work involves diffuse scattering for static magnetic structure factor $S(\mathbf{Q})$, the architecture is well suited for inelastic scattering with dynamical structure factor $S(\mathbf{Q},\omega)$, since the forward model that obtains $S(\mathbf{Q})$ from a parameterized Hamiltonian $H\{J\}$ can also be used to calculate $S(\mathbf{Q},\omega)$. The workflow is shown in **Figure *10***. The Monte Carlo-based forward model is used to compute the $S^{\text{sim}}(\mathbf{Q})$. Instead of directly comparing $S^{\text{sim}}(\mathbf{Q})$ to $S^{\text{exp}}(\mathbf{Q})$ which could suffer from data artifacts, an autoencoder is applied to compress the structure factor $S(\mathbf{Q})$ into a latent-space $L$ where $S_L = \{S_1, S_2, ..., S_D\}$ with $D = 30 < \dim\{\mathbf{Q}\}$. The optimization process of the parameters $\{J\}$ thus happens in the latent space of the autoencoder by comparing $S_L^{\text{exp}}$ and $S_L^{\text{sim}}$. This example demonstrates a generic principle of how machine learning can aid inelastic scattering to probe magnetic orderings and excitations. In particular, if the forward problem of calculating dynamical structure factor $S^{\text{sim}}(\mathbf{Q},\omega)$ from some parameterized Hamiltonian $H\{J\}$ becomes feasible, such as using linear spin-wave theory[188], we expect similar machine learning models will have huge potential to study magnetic excitations with experiment data $S^{\text{exp}}(\mathbf{Q},\omega)$ in strongly correlated systems.

# IV. Experimental Infrastructure and Data

## IV.1. Instrument and beam

Thus far, the discussion has focused on using machine learning-augmented elastic and inelastic scattering, and spectroscopies to better elucidate materials properties. Given the central role of beamline infrastructure in a successful scattering experiment, machine learning has also been applied to optimize instrument operation[189-192]. Li *et al.* achieve the dynamic aperture optimization using machine learning for the storage ring at National Synchrotron Light II (NSLS-II) in Brookhaven National Laboratory (BNL)[189]. Dynamic aperture optimization aims to tune the configuration of the sextupole magnets to increase the ultra-relativistic electron lifetime in the storage ring. It is a multi-objective optimization problem with more than one objective functions $f_m(\mathbf{x}),\ m \geq 2$, to minimize within the parameter space $\{\mathbf{x}\}$, which can be solved by a conventional multi-objective genetic algorithm with further augmentation by machine learning. The direct tracking of a large number of particles forms the "population" in the parameter space $\{\mathbf{x}\}$. The populations in a generic 2D parameter space $(x_1, x_2)$ are shown in **Figure *11*a**. Using *k*-means clustering, the populations are classified as different clusters (**Figure *11*a**, step 1). By evaluating the fitness function, which is the weighted average of objective functions $F(\mathbf{x}) = \sum_{m=1} \omega_m f_m(\mathbf{x})$ and plays the same role as cost function, the populations are further labeled with quality (**Figure *11*a**, step 2), where the best "elite" label corresponds to those that optimize the most objective functions $f_m(\mathbf{x})$ (and have longest electron lifetime in storage ring). Finally, some proportion of candidates among the entire generation are replaced with potentially more competitive candidates repopulated from the regime of the "elite" population (**Figure *11*a**, step 3). The replacement proportion in each intervention can further be dynamically adjusted or skipped based on a discrepancy score evaluated from *k*-nearest neighbor regression and actual fitness function. The use of machine learning accelerates the convergence towards optimized parameters (**Figure *11*b**) and increases the number of high-quality elite candidates to reach longer-term electron beam stability in the storage ring.

In a different example, Leemann *et al.* apply machine learning to study the synchrotron source size stabilization from previous instrumental conditions at Advanced Light Source (ALS) in Lawrence Berkeley National Laboratory (LBL)[190]. The electron beam size can vary (**Figure *11*c**, top) due to the insertion device gaps (**Figure *11*c**, bottom). By constructing a neural-network-based supervised learning model with $x$=Insertion device gaps or phase configurations, $y$=beam size, it can be shown that the neural network outperforms simple regression models and can better capture the beam sizes (**Figure *11*d**, top) with less error (**Figure *11*d,** bottom). It is worth mentioning that the chosen fully-connected artificial neural network contains 3 hidden layers and more parameters, which may also contribute to the superior performance than polynomial regression models.

### IV.2. Data collection and processing

Machine learning can also greatly facilitate the scattering data collection and processing. Here, by "processing" we mean procedures like data refinement, denoising, automatic information-background segmentation etc., but do not include extracting further materials' information. Given the precious beamtime resources, the central question is to extract the same amount information with reduced beamtime. For diffractometry, one typical problem is the diffraction peak-background segmentation, which usually requires fine-collected diffraction spots. Sullivan *et al.* apply a deep U-net to extract the shape of the Bragg peaks from time-of-flight neutron diffraction[82] and X-ray diffraction, which enables a more reliable peak area integration[84]. Training data are augmented with the following operations (**Figure *12*a**):

$$x^E = \text{Histogramming, Rotation, Recenter, Noise, Crop in reciprocal space}$$

In another example, Ke *et al.* apply CNN to identify diffraction spots from noisy data taken from X-ray free-electron laser[83].

For small-angle scattering, given the rapid drop of intensity at high-$Q$ range (for 3D object, $I(Q) \propto Q^{-4}$) and limited beamtime resources, a typical problem lies in optimizing the data collection strategy at the high-$Q$ regime. Asahara *et al.* apply Gaussian mixture modeling to predict longer-time SANS spectra with its prior coming from B-spline regression. The proposed B-spline Gaussian mixture model (BSGMM) outperforms conventional kernel density estimation (KDE) algorithms (**Figure *12*b**)[193] and shortens the SANS experiment by a factor of 5.

Measurements can also be accelerated by reducing necessary sampling points in parameter space with guidance from machine learning. Kanazawa *et al.* propose a workflow that optimizes the automatic sequential $Q$-sampling, which suggests the next $Q$- point based on uncertainties estimated from previously measured data (**Figure *12*c**)[194]. Noack *et al*.[195,196] have used kriging[197,198], a Gaussian process regression method, to design experimental sampling strategy in spatially-resolved SAXS measurement of block copolymer thin films. Compared to a complete set of SAXS measurements sampled using a regular grid as is done traditionally, the authors show that the use of kriging and its variants they have developed in required only a fraction of the sampled spatial coordinates, while arriving at a reconstruction with comparable detail to the outcome of the grid scan. Their closed-loop approach highlights the potential for experimental automation to improve the efficiency in data acquisition and to maximize the information gathered from fragile samples.

Chang *et al.* address a similar challenge by applying CNN to SANS spectra data to reach super-resolution[107]. Even for anisotropic scattering, the CNN-based super-resolution reconstruction allows a better agreement to the ground truth than the conventional bicubic algorithm (**Figure *12*d**).

Finally, machine learning can also be applied in problems that improve other data collection processes, such as calibrating the rotation axis for X-ray tomography[199], improving the phase contrast-spatial resolution contradiction in phase-contrast imaging[200], optimizing data segmentation in transmission X-ray microscopy (TXM)[201], data visualization in neutron scattering data[202], and achieving super-resolution in X-ray tomography[203].

## V. Outlook

### V.1. Machine learning on time-resolved spectroscopies

A wide variety of machine learning models are available to study the dynamics of physical systems, for example, the recurrent neural network (RNN) and RNN-based architectures. They can be used for metamodeling of structural dynamics[204], inferring quantum evolution of superconducting qubits[205], and for modeling quantum many-body systems on large lattices[206]. RNN based models have been applied to study spectra such as nonlinear tomographic absorption spectroscopy[207] and optical spectra for optoelectronic polymers[208]. However, their applications to time-resolved neutron or X-ray scattering are still scarce. In the context of scattering measurements, extra challenges exist given that physical processes are now reflected through neutron or photon counts on detector arrays, accompanied by noise and loss of phase information. Fortunately, neural networks are good at denoising[209], solving phase retrieval problems[210,211], and dealing with information missing in time series[212]. Thus RNN based models can serve as promising techniques to extract deeper insight from time-resolved neutron and X-ray spectra.

Neural ordinary differential equations (Neural ODE) is an alternative framework that can be used to learn from time-series data[213]. This framework can be intimately related to physical models, it is able to make good extrapolation with limited training data and thus has found its applications in quantum phenomena[214,215]. It will become rather interesting to see how neural networks can be combined with physical models and be learned from scattering data with Neural ODE to exploit new physics understanding. Another approach for learning complex nonlinear dynamics is deep Koopman operators[216,217]: where an autoencoder-like structure is developed to connect observed states with intrinsic states represented by the learned Koopman coordinates, where the intrinsic states get evolved with learned dynamics within the latent space. Such an architecture can be analogously mapped to physical observables, i.e., scattering data, and the intrinsic quantum states of measured specimens, thus can also serve as a promising approach to interpret time-resolved scattering data.

### V.2. Leveraging information in real and reciprocal spaces

Frameworks that employ the principles of symmetry and Fourier transforms could efficiently learn models of complex physical systems and help us efficiently harness scattering data, in either real space, reciprocal space, or both.

Symmetry and Fourier transforms are two of the most valuable and commonly used computational tools for tackling complex physics problems. These tools encode much of the domain knowledge we have about arbitrary scientific data in 3D space: 1) The properties of physical systems (geometry and geometric tensor fields) transform predictably under rotations, translations, and inversion (aka, the 3D Euclidean symmetry). 2) While physical systems can be described with equal accuracy in both real (position) space and reciprocal (momentum) space, some patterns/operations (e.g., convolutions, derivatives) are much simpler to identify/evaluate in one space than the other. The beauty of symmetry and Fourier transforms is that they make no assumptions about the incoming data (only that it exists in 3D Euclidean space); this generality is also an opportunity for improvement. The strength of machine learning is the ability to build efficient algorithms by leveraging the context contained in a given dataset to forgo expensive computation.

A constant theme in scattering data is acquiring data in reciprocal space and having that data to inform something traditionally represented in real space. While there are models that can operate on these domains separately, it would be a valuable and natural direction to extend these methods to simultaneously operate and exchange information in both spaces. This would also allow the user to input and output data in whichever space is more convenient and intuitive, and can directly support methods like diffraction imaging, which contain information in both spaces.

Using learnable context in combination with the fundamental principles of symmetry and Fourier transforms could help alleviate some of the primary challenges associated with scattering experiments: missing phase information and sampling. Additionally, frameworks that can simultaneously compute in and exchange information between real and reciprocal space could naturally predict quasiparticle bandstructures from real space coordinates and express charge densities in terms of commonly used plane-wave basis sets.

### V.3. Multimodal machine learning

Materials characterization often requires insight from multiple experimental techniques with sensitivity to different types of excitations in order to gain a complete understanding of the properties and behaviors of materials. Data acquired using different neutron and X-ray scattering techniques are oftentimes complementary but are typically synthesized manually by researchers. In this regard, machine learning may provide an important avenue toward intelligent analysis across multiple modalities. Multimodal machine learning[218-221] has already been explored for a range of versatile applications, including activity and context detection[222,223]; recognition of objects[224], images[225], and emotions[226]; and improving certain medical diagnostics[227,228]. By consolidating information from multiple, complementary sources, multimodal machine learning models have the potential to make more robust predictions and discover more sophisticated relationships between data. At the same time, this approach introduces new prerequisites compared to learning from single modalities. The taxonomy by Baltrušaitis *et al.* considers five principal challenges of multimodal machine learning[220]: 1) representation of heterogeneous data, 2) translation, or mapping, of data from one modality to another, 3) alignment between elements of two or more different modalities, 4) fusion of information to perform a prediction, and 5) co-learning, which considers how knowledge gained by learning from one modality can assist a model trained on a different modality whose resources may be more limited. These are likewise important considerations for the application of multimodal machine learning in the context of neutron and X-ray data analysis: Different experimental techniques access widely different energy, time, length, and momentum scales, produce diverse data structures, and carry varying levels of uncertainty. Additionally, developing the data infrastructure to aggregate measurements from multiple instruments would be an important undertaking for neutron and X-ray facilities as a whole. Nonetheless, intelligent synthesis of multiple experimental signatures appears to be a promising direction to better extract insights from data and possibly accelerate materials design and discovery.

### V.4. High-performance computing for quantum materials

Increasingly, studies in functional materials underscore quantum phenomena emergent from entanglement. These quantum phenomena, such as quantum spin liquids, unconventional superconductivity, and many-body localization, are beyond the structure information description. However, the associated correlations are encoded in the inelastic scattering spectroscopies through the energy-momentum resolution, which motivates corresponding theoretical predictions. Due to quantum entanglement, semiclassical theories like the mean-field theory, linear spin-wave theory, or even DFT become insufficient due to the absence of static or dynamic electron correlations. Thus, the machine learning and cross-validation of spectroscopies associated with these materials require sophisticated computational methods.

To sufficiently include the quantum entanglement in spectral calculations, two promising routes have been widely attempted. The first route is the correction of DFT by embedding other methods. Beyond the elementary DFT+U corrections for total energy, the GW method allows a self-consistent correction of the Green's function using the screened Coulomb interaction in the random-phase approximation (RPA) form[229]. A more sophisticated correction for strong correlation effects is the DFT + DMFT (dynamical mean-field theory) method[230]. By mapping the self-energy into a single-site impurity problem, DMFT further includes local high-order correlations in the spectral calculations[231]. These corrections on top of DFT enable spectral calculations for materials with substantial quantum entanglement. However, as the corrections are usually biased, the accuracy of the results is sometimes not well-controlled. The DFT+DMFT method has been widely used to simulate the single-particle Green's function relevant for photoemission experiments[232]. Its numerical complexity increases dramatically when extended to two-particle or four-particle correlation functions, which are required to evaluate inelastic scattering cross-sections. Implemented using the Bethe-Salpeter equation and Lanczos method, the DFT+DMFT methods have been recently applied to the simulation of neutron scattering and resonant inelastic X-ray scattering (RIXS) spectra[233,234], correctly reflecting the multiplet effects and Mott transition in transition-metal materials.

The other route is constructing effective low-energy models based on the *ab initio* Wannier orbitals and evaluate spectral properties based on this highly-entangled effective model. Along this route, wavefunction-based methods including exact diagonalization[235], coupled clusters[236], and density-matrix renormalization group (DMRG)[237] provide exact or asymptotically exact solutions to excited-state spectra for arbitrarily strong correlations. The disadvantage of these wavefunction-based methods is that the rapid scaling of the computational complexity restricts the calculations to only small systems or low dimension with limited numbers of bands. Another class of the model-based unbiased methods is the quantum Monte Carlo[238], which is less sensitive to the system's size but restricted by high temperature. These methods have been widely used in the scattering spectral calculations for spin liquids[239] and unconventional superconductors[240], where spin correlations are dominated in a few bands.

Spectral calculations based on either route are computationally expensive and require massively parallel computing techniques. Most methods exhibit good scaling performance in distributed computing. With the reconstruction of bottom-level linear algebraic operations, these approaches have been further accelerated using the General-Purpose Graphics Processing Unit (GPGPU). In addition to providing a high-throughput dataset for machine learning, recent studies have demonstrated that machine learning can also benefit these numerical calculations by improving the efficiency and accuracy[241,242].

**Acknowledgments**

Z.C., N.A. and M.L. acknowledges support from U.S. DOE BES Award No. DE-SC0020148. N.A. acknowledges National Science Foundation (NSF) GRFP support under Grant No. 1122374. Y.W. acknowledges support from NSF award DMR-2038011. R.P.X. and R.E. acknowledge the support by BiGmax, the Max Planck Society's Research Network on Big-Data-Driven Materials-Science, the European Research Council (ERC) under the European Union's Horizon 2020 research and innovation program Grant No. ERC-2015-CoG-682843.

**References**

1 Lovesey, S. W. *Theory of neutron scattering from condensed matter*. (Clarendon Press, 1984).
2 Als-Nielsen, J. & McMorrow, D. *Elements of modern X-ray physics*. 2nd edn, (Wiley, 2011).
3 Bohn, F. *et al.* European source of science. *The ESS Project* **1** (2002).
4 National Research Council (U.S.). Committee on Atomic Molecular and Optical Sciences 2010. *Controlling the quantum world : the science of atoms, molecules, and photons*. (National Academies Press, 2007).
5 Yabashi, M. & Tanaka, H. The next ten years of X-ray science. *Nature Photonics* **11**, 12-14, doi:10.1038/nphoton.2016.251 (2017).
6 Su, X. D. *et al.* Protein Crystallography from the Perspective of Technology Developments. *Crystallogr Rev* **21**, 122-153, doi:10.1080/0889311X.2014.973868 (2015).
7 Shpyrko, O. G. X-ray photon correlation spectroscopy. *J Synchrotron Radiat* **21**, 1057-1064, doi:10.1107/S1600577514018232 (2014).
8 Murphy, K. P. *Machine learning : a probabilistic perspective*. (MIT Press, 2012).
9 Goodfellow, I., Bengio, Y. & Courville, A. *Deep learning*. (The MIT Press, 2016).
10 Brunton, S. L., Proctor, J. L. & Kutz, J. N. Discovering governing equations from data by sparse identification of nonlinear dynamical systems. *Proceedings of the National Academy of Sciences* **113**, 3932, doi:10.1073/pnas.1517384113 (2016).

11 Kaheman, K., Kaiser, E., Strom, B., Kutz, J. N. & Brunton, S. L. Learning discrepancy models from experimental data. *arXiv preprint arXiv:1909.08574* (2019).

12 Rackauckas, C. *et al.* Universal differential equations for scientific machine learning. *arXiv preprint arXiv:2001.04385* (2020).

13 Niu, M. Y., Horesh, L. & Chuang, I. Recurrent neural networks in the eye of differential equations. *arXiv preprint arXiv:1904.12933* (2019).

14 Sherstinsky, A. Fundamentals of Recurrent Neural Network (RNN) and Long Short-Term Memory (LSTM) network. *Physica D: Nonlinear Phenomena* **404**, 132306, doi:10.1016/j.physd.2019.132306 (2020).

15 De Silva, B. M., Higdon, D. M., Brunton, S. L. & Kutz, J. N. Discovery of Physics From Data: Universal Laws and Discrepancies. *Frontiers in Artificial Intelligence* **3**, doi:10.3389/frai.2020.00025 (2020).

16 Schütt, K. T. *et al.* How to represent crystal structures for machine learning: Towards fast prediction of electronic properties. *Physical Review B* **89**, doi:10.1103/PhysRevB.89.205118 (2014).

17 Schütt, K. T. *et al.* SchNet: A continuous-filter convolutional neural network for modeling quantum interactions. *arXiv e-prints*, arXiv:1706.08566 (2017).

18 Schütt, K. T., Sauceda, H. E., Kindermans, P.-J., Tkatchenko, A. & Müller, K.-R. SchNet – A deep learning architecture for molecules and materials. *The Journal of Chemical Physics* **148**, 241722, doi:10.1063/1.5019779 (2018).

19 Nathaniel Thomas, T. S., Steven Kearnes, Lusann Yang, Li Li, Kai Kohlhoff, Patrick Riley. Tensor field networks: Rotation- and translation-equivariant neural networks for 3D point clouds. *arXiv preprint* (2018).

20 Miller, B. K., Geiger, M., Smidt, T. E. & Noé, F. Relevance of rotationally equivariant convolutions for predicting molecular properties. *arXiv preprint arXiv:2008.08461* (2020).

21 Schütt, K. T., Arbabzadah, F., Chmiela, S., Müller, K. R. & Tkatchenko, A. Quantum-chemical insights from deep tensor neural networks. *Nature Communications* **8**, 13890, doi:10.1038/ncomms13890 (2017).

22 Schütt, K. T., Gastegger, M., Tkatchenko, A., Müller, K. R. & Maurer, R. J. Unifying machine learning and quantum chemistry with a deep neural network for molecular wavefunctions. *Nature Communications* **10**, 5024, doi:10.1038/s41467-019-12875-2 (2019).

23 Pun, G. P. P., Batra, R., Ramprasad, R. & Mishin, Y. Physically informed artificial neural networks for atomistic modeling of materials. *Nature Communications* **10**, 2339, doi:10.1038/s41467-019-10343-5 (2019).

24 Xie, T. & Grossman, J. C. Crystal Graph Convolutional Neural Networks for an Accurate and Interpretable Prediction of Material Properties. *Phys Rev Lett* **120**, 145301, doi:10.1103/PhysRevLett.120.145301 (2018).

25 Xie, T., France-Lanord, A., Wang, Y., Shao-Horn, Y. & Grossman, J. C. Graph dynamical networks for unsupervised learning of atomic scale dynamics in materials. *Nature Communications* **10**, 2667, doi:10.1038/s41467-019-10663-6 (2019).

26 Chen, C., Ye, W., Zuo, Y., Zheng, C. & Ong, S. P. Graph Networks as a Universal Machine Learning Framework for Molecules and Crystals. *Chemistry of Materials* **31**, 3564-3572, doi:10.1021/acs.chemmater.9b01294 (2019).

27 Isayev, O. *et al.* Universal fragment descriptors for predicting properties of inorganic crystals. *Nat Commun* **8**, 15679, doi:10.1038/ncomms15679 (2017).

28 Pilania, G., Wang, C., Jiang, X., Rajasekaran, S. & Ramprasad, R. Accelerating materials property predictions using machine learning. *Sci Rep* **3**, 2810, doi:10.1038/srep02810 (2013).

29 Carrete, J., Li, W., Mingo, N., Wang, S. & Curtarolo, S. Finding Unprecedentedly Low-Thermal-Conductivity Half-Heusler Semiconductors via High-Throughput Materials Modeling. *Physical Review X* **4**, doi:10.1103/PhysRevX.4.011019 (2014).

30 Tawfik, S. A., Isayev, O., Spencer, M. J. S. & Winkler, D. A. Predicting Thermal Properties of Crystals Using Machine Learning. *Advanced Theory and Simulations* **3**, doi:10.1002/adts.201900208 (2019).

31 van Roekeghem, A., Carrete, J., Oses, C., Curtarolo, S. & Mingo, N. High-Throughput Computation of Thermal Conductivity of High-Temperature Solid Phases: The Case of Oxide and Fluoride Perovskites. *Physical Review X* **6**, doi:10.1103/PhysRevX.6.041061 (2016).

32 Dong, Y. *et al.* Bandgap prediction by deep learning in configurationally hybridized graphene and boron nitride. *npj Computational Materials* **5**, doi:10.1038/s41524-019-0165-4 (2019).

33 Meredig, B. *et al.* Can machine learning identify the next high-temperature superconductor? Examining extrapolation performance for materials discovery. *Molecular Systems Design & Engineering* **3**, 819-825, doi:10.1039/c8me00012c (2018).

34 Scheurer, M. S. & Slager, R. J. Unsupervised Machine Learning and Band Topology. *Phys Rev Lett* **124**, 226401, doi:10.1103/PhysRevLett.124.226401 (2020).

35 Stanev, V. *et al.* Machine learning modeling of superconducting critical temperature. *npj Computational Materials* **4**, doi:10.1038/s41524-018-0085-8 (2018).

36 Ward, L., Agrawal, A., Choudhary, A. & Wolverton, C. A general-purpose machine learning framework for predicting properties of inorganic materials. *npj Computational Materials* **2**, doi:10.1038/npjcompumats.2016.28 (2016).

37 Zhuo, Y., Mansouri Tehrani, A. & Brgoch, J. Predicting the Band Gaps of Inorganic Solids by Machine Learning. *J Phys Chem Lett* **9**, 1668-1673, doi:10.1021/acs.jpclett.8b00124 (2018).

38 Nina Andrejevic, J. A., Chris H. Rycroft, Mingda Li. Machine learning spectral indicators of topology. *arXiv preprint* (2020).

39 Gomez-Bombarelli, R. *et al.* Automatic Chemical Design Using a Data-Driven Continuous Representation of Molecules. *ACS Cent Sci* **4**, 268-276, doi:10.1021/acscentsci.7b00572 (2018).

40 Liu, Y., Zhao, T., Ju, W. & Shi, S. Materials discovery and design using machine learning. *Journal of Materiomics* **3**, 159-177, doi:10.1016/j.jmat.2017.08.002 (2017).

41 Oliynyk, A. O. *et al.* High-Throughput Machine-Learning-Driven Synthesis of Full-Heusler Compounds. *Chemistry of Materials* **28**, 7324-7331, doi:10.1021/acs.chemmater.6b02724 (2016).

42 Raccuglia, P. *et al.* Machine-learning-assisted materials discovery using failed experiments. *Nature* **533**, 73-76, doi:10.1038/nature17439 (2016).

43 Noh, J. *et al.* Inverse Design of Solid-State Materials via a Continuous Representation. *Matter* **1**, 1370-1384, doi:10.1016/j.matt.2019.08.017 (2019).

44 Kim, S., Noh, J., Gu, G. H., Aspuru-Guzik, A. & Jung, Y. Generative Adversarial Networks for Crystal Structure Prediction. *ACS Cent Sci* **6**, 1412-1420, doi:10.1021/acscentsci.0c00426 (2020).

45 Botu, V., Batra, R., Chapman, J. & Ramprasad, R. Machine Learning Force Fields: Construction, Validation, and Outlook. *The Journal of Physical Chemistry C* **121**, 511-522, doi:10.1021/acs.jpcc.6b10908 (2016).

46 Glielmo, A., Sollich, P. & De Vita, A. Accurate interatomic force fields via machine learning with covariant kernels. *Physical Review B* **95**, doi:10.1103/PhysRevB.95.214302 (2017).

47 Kruglov, I., Sergeev, O., Yanilkin, A. & Oganov, A. R. Energy-free machine learning force field for aluminum. *Sci Rep* **7**, 8512, doi:10.1038/s41598-017-08455-3 (2017).

48 Li, Z., Kermode, J. R. & De Vita, A. Molecular dynamics with on-the-fly machine learning of quantum-mechanical forces. *Phys Rev Lett* **114**, 096405, doi:10.1103/PhysRevLett.114.096405 (2015).

49 Zhang, L., Lin, D.-Y., Wang, H., Car, R. & E, W. Active learning of uniformly accurate interatomic potentials for materials simulation. *Physical Review Materials* **3**, doi:10.1103/PhysRevMaterials.3.023804 (2019).

50 Mortazavi, B. *et al.* Machine-learning interatomic potentials enable first-principles multiscale modeling of lattice thermal conductivity in graphene/borophene heterostructures. *Materials Horizons* **7**, 2359-2367, doi:10.1039/d0mh00787k (2020).

51 Deringer, V. L. *et al.* Origins of structural and electronic transitions in disordered silicon. *Nature* **589**, 59-64, doi:10.1038/s41586-020-03072-z (2021).

52 Butler, K. T., Davies, D. W., Cartwright, H., Isayev, O. & Walsh, A. Machine learning for molecular and materials science. *Nature* **559**, 547-555, doi:10.1038/s41586-018-0337-2 (2018).

53 Rupp, M. Machine learning for quantum mechanics in a nutshell. *Int J Quantum Chem* **115**, 1058-1073, doi:10.1002/qua.24954 (2015).

54 Schmidt, J., Marques, M. R. G., Botti, S. & Marques, M. A. L. Recent advances and applications of machine learning in solid-state materials science. *npj Computational Materials* **5**, doi:10.1038/s41524-019-0221-0 (2019).

55 Mehta, P. *et al.* A high-bias, low-variance introduction to Machine Learning for physicists. *Phys Rep* **810**, 1-124, doi:10.1016/j.physrep.2019.03.001 (2019).

56 Carleo, G. *et al.* Machine learning and the physical sciences. *Rev Mod Phys* **91**, doi:10.1103/revmodphys.91.045002 (2019).

57 Batra, R., Song, L. & Ramprasad, R. Emerging materials intelligence ecosystems propelled by machine learning. *Nature Reviews Materials*, doi:10.1038/s41578-020-00255-y (2020).

58 Suh, C., Fare, C., Warren, J. A. & Pyzer-Knapp, E. O. Evolving the Materials Genome: How Machine Learning Is Fueling the Next Generation of Materials Discovery. *Annual Review of Materials Research* **50**, 1-25, doi:10.1146/annurev-matsci-082019-105100 (2020).

59 Wolpert, D. H. & Macready, W. G. No free lunch theorems for optimization. *IEEE Transactions on Evolutionary Computation* **1**, 67-82, doi:10.1109/4235.585893 (1997).

60 Rupp, M., Tkatchenko, A., Muller, K. R. & von Lilienfeld, O. A. Fast and accurate modeling of molecular atomization energies with machine learning. *Phys Rev Lett* **108**, 058301, doi:10.1103/PhysRevLett.108.058301 (2012).

61 Faber, F., Lindmaa, A., von Lilienfeld, O. A. & Armiento, R. Crystal structure representations for machine learning models of formation energies. *International Journal of Quantum Chemistry* **115**, 1094-1101, doi:10.1002/qua.24917 (2015).

62 Behler, J. & Parrinello, M. Generalized neural-network representation of high-dimensional potential-energy surfaces. *Phys Rev Lett* **98**, 146401, doi:10.1103/PhysRevLett.98.146401 (2007).

63 Bartók, A. P., Kondor, R. & Csányi, G. On representing chemical environments. *Physical Review B* **87**, doi:10.1103/PhysRevB.87.184115 (2013).

64 Hoffmann, J. *et al.* Data-driven approach to encoding and decoding 3-D crystal structures. *arXiv preprint arXiv:1909.00949* (2019).

65 Ronneberger, O., Fischer, P. & Brox, T. in *International Conference on Medical image computing and computer-assisted intervention.* 234--241 (Springer).

66 Kingma, D. P. & Welling, M. Auto-encoding variational bayes. *arXiv preprint arXiv:1312.6114* (2013).

67 Samarakoon, A. M. *et al.* Machine-learning-assisted insight into spin ice Dy2Ti2O7. *Nat Commun* **11**, 892, doi:10.1038/s41467-020-14660-y (2020).

68 Goodfellow, I. J. *et al.* in *Proceedings of the International Conference on Neural Information Processing Systems* 2672–2680.

69 Maurice Weiler, M. G., Max Welling, Wouter Boomsma, Taco Cohen. 3D Steerable CNNs: Learning Rotationally Equivariant Features in Volumetric Data. *32nd Conference on Neural Information Processing Systems (NeurIPS 2018)* (2018).

70 Risi Kondor, Z. L., Shubhendu Trivedi. Clebsch–Gordan Nets: a Fully Fourier Space Spherical Convolutional Neural Network. *32nd Conference on Neural Information Processing Systems (NeurIPS 2018)* (2018).

71 Paatero, P. & Tapper, U. Positive matrix factorization: A non-negative factor model with optimal utilization of error estimates of data values. *Environmetrics* **5**, 111-126, doi:https://doi.org/10.1002/env.3170050203 (1994).

72 Lee, D. D. & Seung, H. S. Learning the parts of objects by non-negative matrix factorization. *Nature* **401**, 788-791, doi:10.1038/44565 (1999).

73 Kitamura, D. *et al.* in *IEEE International Symposium on Signal Processing and Information Technology.* 000392--000397.

74 Kim, G. *et al.* First-principles and machine learning predictions of elasticity in severely lattice-distorted high-entropy alloys with experimental validation. *Acta Materialia* **181**, 124-138, doi:10.1016/j.actamat.2019.09.026 (2019).

75 Garcia-Cardona, C. *et al.* Learning to Predict Material Structure from Neutron Scattering Data. *Ieee Int Conf Big Da*, 4490-4497 (2019).
76 Oviedo, F. *et al.* Fast and interpretable classification of small X-ray diffraction datasets using data augmentation and deep neural networks. *npj Computational Materials* **5**, doi:10.1038/s41524-019-0196-x (2019).
77 Liu, C. H., Tao, Y. Z., Hsu, D., Du, Q. & Billinge, S. J. L. Using a machine learning approach to determine the space group of a structure from the atomic pair distribution function. *Acta Crystallogr A* **75**, 633-643, doi:10.1107/S2053273319005606 (2019).
78 Olds, D. *et al.* Precise implications for real-space pair distribution function modeling of effects intrinsic to modern time-of-flight neutron diffractometers. *Acta Crystallogr A Found Adv* **74**, 293-307, doi:10.1107/S2053273318003224 (2018).
79 Long, C. J., Bunker, D., Li, X., Karen, V. L. & Takeuchi, I. Rapid identification of structural phases in combinatorial thin-film libraries using x-ray diffraction and non-negative matrix factorization. *Rev Sci Instrum* **80**, 103902, doi:10.1063/1.3216809 (2009).
80 Stanev, V. *et al.* Unsupervised phase mapping of X-ray diffraction data by nonnegative matrix factorization integrated with custom clustering. *npj Computational Materials* **4**, doi:10.1038/s41524-018-0099-2 (2018).
81 Venderley, J. *et al.* Harnessing Interpretable and Unsupervised Machine Learning to Address Big Data from Modern X-ray Diffraction. *arXiv pre-print server,* doi:arxiv:2008.03275 (2020).
82 Sullivan, B. *et al.* Volumetric Segmentation via Neural Networks Improves Neutron Crystallography Data Analysis. *Ieee Acm Int Symp*, 549-555, doi:10.1109/Ccgrid.2019.00070 (2019).
83 Ke, T. W. *et al.* A convolutional neural network-based screening tool for X-ray serial crystallography. *J Synchrotron Radiat* **25**, 655-670, doi:10.1107/S1600577518004873 (2018).
84 Sullivan, B. *et al.* BraggNet: integrating Bragg peaks using neural networks. *Journal of Applied Crystallography* **52**, 854-863, doi:10.1107/S1600576719008665 (2019).
85 Song, Y. T., Tamura, N., Zhang, C. B., Karami, M. & Chen, X. Data-driven approach for synchrotron X-ray Laue microdiffraction scan analysis. *Acta Crystallogr A* **75**, 876-888, doi:10.1107/S2053273319012804 (2019).
86 Rodman, J., Lin, Y. W., Sprouster, D., Ecker, L. & Yoo, S. Automated X-Ray Diffraction of Irradiated Materials. *2017 New York Scientific Data Summit (Nysds)* (2017).
87 Glatter, O. & Kratky, O. *Small angle x-ray scattering*. (Academic Press, 1982).
88 Svergun, D. I., Feĭgin, L. A. & Taylor, G. W. *Structure analysis by small-angle x-ray and neutron scattering*. (Plenum Press, 1987).
89 Sinha, S. K., Sirota, E. B., Garoff, S. & Stanley, H. B. X-Ray and Neutron-Scattering from Rough Surfaces. *Physical Review B* **38**, 2297-2311, doi:DOI 10.1103/PhysRevB.38.2297 (1988).

90 Pedersen, J. S. Analysis of small-angle scattering data from colloids and polymer solutions: modeling and least-squares fitting. *Advances in Colloid and Interface Science* **70**, 171-210, doi:https://doi.org/10.1016/S0001-8686(97)00312-6 (1997).

91 Chu, B. & Hsiao, B. S. Small-angle X-ray scattering of polymers. *Chem Rev* **101**, 1727-1761, doi:10.1021/cr9900376 (2001).

92 Roe, R. J. *Methods of X-ray and neutron scattering in polymer science*. (Oxford University Press, 2000).

93 Koch, M. H. J., Vachette, P. & Svergun, D. I. Small-angle scattering: a view on the properties, structures and structural changes of biological macromolecules in solution. *Q Rev Biophys* **36**, 147-227, doi:10.1017/S0033583503003871 (2003).

94 Ashkar, R. *et al.* Neutron scattering in the biological sciences: progress and prospects. *Acta Crystallogr D Struct Biol* **74**, 1129-1168, doi:10.1107/S2059798318017503 (2018).

95 Lattman, E. E., Grant, T. D. & Snell, E. H. *Biological small angle scattering : theory and practice*. First edition. edn, (Oxford University Press, 2018).

96 Mühlbauer, S. *et al.* Magnetic small-angle neutron scattering. *Rev Mod Phys* **91**, doi:10.1103/RevModPhys.91.015004 (2019).

97 Das, P. *et al.* Small-angle neutron scattering study of vortices in superconducting Ba(Fe0.93Co0.07)2As2. *Superconductor Science and Technology* **23**, 054007, doi:10.1088/0953-2048/23/5/054007 (2010).

98 Eskildsen, M. R., Forgan, E. M. & Kawano-Furukawa, H. Vortex structures, penetration depth and pairing in iron-based superconductors studied by small-angle neutron scattering. *Rep Prog Phys* **74**, 124504, doi:10.1088/0034-4885/74/12/124504 (2011).

99 Li, Y., Egetenmeyer, N., Gavilano, J. L., Barišić, N. & Greven, M. Magnetic vortex lattice in HgBa2CuO4+δobserved by small-angle neutron scattering. *Physical Review B* **83**, doi:10.1103/PhysRevB.83.054507 (2011).

100 Lee, E. Y., Fulan, B. M., Wong, G. C. L. & Ferguson, A. L. Mapping membrane activity in undiscovered peptide sequence space using machine learning. *P Natl Acad Sci USA* **113**, 13588-13593, doi:10.1073/pnas.1609893113 (2016).

101 Wang, B. Y., Yager, K., Yu, D. T. & Hoai, M. X-ray Scattering Image Classification Using Deep Learning. *Ieee Wint Conf Appl*, 697-704, doi:10.1109/Wacv.2017.83 (2017).

102 Franke, D., Jeffries, C. M. & Svergun, D. I. Machine Learning Methods for X-Ray Scattering Data Analysis from Biomacromolecular Solutions. *Biophys J* **114**, 2485-2492, doi:10.1016/j.bpj.2018.04.018 (2018).

103 Demerdash, O. *et al.* Using Small-Angle Scattering Data and Parametric Machine Learning to Optimize Force Field Parameters for Intrinsically Disordered Proteins. *Front Mol Biosci* **6**, 64, doi:10.3389/fmolb.2019.00064 (2019).

104 Hura, G. L. *et al.* Small angle X-ray scattering-assisted protein structure prediction in CASP13 and emergence of solution structure differences. *Proteins* **87**, 1298-1314, doi:10.1002/prot.25827 (2019).

105 Liu, S. *et al.* Convolutional neural networks for grazing incidence x-ray scattering patterns: thin film structure identification. *MRS Communications* **9**, 586-592, doi:10.1557/mrc.2019.26 (2019).

106 Archibald, R. K. *et al.* Classifying and analyzing small-angle scattering data using weighted k nearest neighbors machine learning techniques. *Journal of Applied Crystallography* **53**, 326-334, doi:10.1107/s1600576720000552 (2020).

107 Chang, M.-C., Wei, Y., Chen, W.-R. & Do, C. Deep learning-based super-resolution for small-angle neutron scattering data: attempt to accelerate experimental workflow. *MRS Communications* **10**, 11-17, doi:10.1557/mrc.2019.166 (2020).

108 Chen, Y. L. & Pollack, L. Machine learning deciphers structural features of RNA duplexes measured with solution X-ray scattering. *IUCrJ* **7**, 870-880, doi:10.1107/S2052252520008830 (2020).

109 Do, C., Chen, W.-R. & Lee, S. Small Angle Scattering Data Analysis Assisted by Machine Learning Methods. *MRS Advances* **5**, 1577-1584, doi:10.1557/adv.2020.130 (2020).

110 He, H., Liu, C. & Liu, H. Model Reconstruction from Small-Angle X-Ray Scattering Data Using Deep Learning Methods. *iScience* **23**, 100906, doi:10.1016/j.isci.2020.100906 (2020).

111 Ikemoto, H. *et al.* Classification of grazing-incidence small-angle X-ray scattering patterns by convolutional neural network. *J Synchrotron Radiat* **27**, 1069-1073, doi:10.1107/s1600577520005767 (2020).

112 Lyoussi, A. *et al.* Deep Learning Methods On Neutron Scattering Data. *EPJ Web of Conferences* **225**, 01004, doi:10.1051/epjconf/202022501004 (2020).

113 Kardjilov, N., Manke, I., Hilger, A., Strobl, M. & Banhart, J. Neutron imaging in materials science. *Materials Today* **14**, 248-256, doi:10.1016/s1369-7021(11)70139-0 (2011).

114 Kardjilov, N., Manke, I., Woracek, R., Hilger, A. & Banhart, J. Advances in neutron imaging. *Materials Today* **21**, 652-672, doi:10.1016/j.mattod.2018.03.001 (2018).

115 Jacobsen, C. J. *X-ray microscopy*. (Cambridge University Press, 2020).

116 Erickson, B. J., Korfiatis, P., Akkus, Z. & Kline, T. L. Machine Learning for Medical Imaging. *RadioGraphics* **37**, 505-515, doi:10.1148/rg.2017160130 (2017).

117 Wang, G., Ye, J. C. & De Man, B. Deep learning for tomographic image reconstruction. *Nature Machine Intelligence* **2**, 737-748, doi:10.1038/s42256-020-00273-z (2020).

118 Pelt, D. M. & Batenburg, K. J. Fast Tomographic Reconstruction From Limited Data Using Artificial Neural Networks. *IEEE Transactions on Image Processing* **22**, 5238-5251, doi:10.1109/tip.2013.2283142 (2013).

119 Micieli, D., Minniti, T., Evans, L. M. & Gorini, G. Accelerating Neutron Tomography experiments through Artificial Neural Network based reconstruction. *Sci Rep-Uk* **9**, doi:10.1038/s41598-019-38903-1 (2019).

120 Yang, X. *et al.* Tomographic reconstruction with a generative adversarial network. *J Synchrotron Radiat* **27**, 486-493, doi:10.1107/S1600577520000831 (2020).

121 Zhang, X. G., Xu, J. J. & Ge, G. Y. Defects recognition on X-ray images for weld inspection using SVM. *Proceedings of the 2004 International Conference on Machine Learning and Cybernetics, Vols 1-7*, 3721-3725 (2004).

122 Rale, A. P., Gharpure, D. C. & Ravindran, V. R. Comparison of different ANN techniques for automatic defect detection in X-Ray Images. *2009 International Conference on Emerging Trends in Electronic and Photonic Devices and Systems (Electro-2009)*, 193-+ (2009).

123 Rovinelli, A., Sangid, M. D., Proudhon, H. & Ludwig, W. Using machine learning and a data-driven approach to identify the small fatigue crack driving force in polycrystalline materials. *npj Computational Materials* **4**, doi:10.1038/s41524-018-0094-7 (2018).

124 Guntoro, P. I., Tiu, G., Ghorbani, Y., Lund, C. & Rosenkranz, J. Application of machine learning techniques in mineral phase segmentation for X-ray microcomputed tomography (μCT) data. *Minerals Engineering* **142**, 105882, doi:10.1016/j.mineng.2019.105882 (2019).

125 Parkinson, D. Y. *et al.* Machine Learning for Micro-Tomography. *Proc Spie* **10391**, doi:10.1117/12.2274731 (2017).

126 Earles, J. M. *et al.* In vivo quantification of plant starch reserves at micrometer resolution using X-ray microCT imaging and machine learning. *New Phytol*, doi:10.1111/nph.15068 (2018).

127 Abidin, A. Z. *et al.* Deep transfer learning for characterizing chondrocyte patterns in phase contrast X-Ray computed tomography images of the human patellar cartilage. *Comput Biol Med* **95**, 24-33, doi:10.1016/j.compbiomed.2018.01.008 (2018).

128 Beliaev, M. *et al.* Quantification of sheet nacre morphogenesis using X-ray nanotomography and deep learning. *J Struct Biol* **209**, 107432, doi:10.1016/j.jsb.2019.107432 (2020).

129 Miao, J., Charalambous, P., Kirz, J. & Sayre, D. Extending the methodology of X-ray crystallography to allow imaging of micrometre-sized non-crystalline specimens. *Nature* **400**, 342-344, doi:10.1038/22498 (1999).

130 Miao, J., Ishikawa, T., Robinson, I. K. & Murnane, M. M. Beyond crystallography: Diffractive imaging using coherent x-ray light sources. *Science* **348**, 530-535, doi:10.1126/science.aaa1394 (2015).

131 Pfeiffer, F. X-ray ptychography. *Nature Photonics* **12**, 9-17, doi:10.1038/s41566-017-0072-5 (2017).

132 Langbehn, B. *et al.* Three-Dimensional Shapes of Spinning Helium Nanodroplets. *Phys Rev Lett* **121**, 255301, doi:10.1103/PhysRevLett.121.255301 (2018).

133 Zimmermann, J. *et al.* Deep neural networks for classifying complex features in diffraction images. *Phys Rev E* **99**, 063309, doi:10.1103/PhysRevE.99.063309 (2019).

134 Cherukara, M. J. *et al.* AI-enabled high-resolution scanning coherent diffraction imaging. *Appl. Phys. Lett.* **117**, 044103, doi:10.1063/5.0013065 (2019).

135 Scheinker, A. & Pokharel, R. Adaptive 3D convolutional neural network-based reconstruction method for 3D coherent diffraction imaging. *Journal of Applied Physics* **128**, 184901, doi:10.1063/5.0014725 (2020).

136 Du, M., Nashed, Y. S. G., Kandel, S., Gursoy, D. & Jacobsen, C. Three dimensions, two microscopes, one code: Automatic differentiation for x-ray nanotomography beyond the depth of focus limit. *Science Advances* **6**, doi:ARTN eaay3700,10.1126/sciadv.aay3700 (2020).

137 Ongie, G. *et al.* Deep Learning Techniques for Inverse Problems in Imaging. *IEEE Journal on Selected Areas in Information Theory* **1**, 39-56, doi:10.1109/jsait.2020.2991563 (2020).

138 Bressler, C. & Chergui, M. Ultrafast X-ray Absorption Spectroscopy. *Chemical Reviews* **104**, 1781-1812, doi:10.1021/cr0206667 (2004).

139 Saisho, H. & Gohshi, Y. *Applications of synchrotron radiation to materials analysis*. (Elsevier, 1996).

140 Rehr, J. J. & Albers, R. C. Theoretical approaches to x-ray absorption fine structure. *Reviews of modern physics* **72**, 621 (2000).

141 Rehr, J. J. *et al.* Ab initio theory and calculations of X-ray spectra. *Comptes Rendus Physique* **10**, 548-559, doi:10.1016/j.crhy.2008.08.004 (2009).

142 Rehr, J. J., Kas, J. J., Vila, F. D., Prange, M. P. & Jorissen, K. Parameter-free calculations of X-ray spectra with FEFF9. *Phys Chem Chem Phys* **12**, 5503-5513, doi:10.1039/B926434E (2010).

143 Carbone, M. R., Yoo, S., Topsakal, M. & Lu, D. Classification of local chemical environments from x-ray absorption spectra using supervised machine learning. *Physical Review Materials* **3**, doi:10.1103/PhysRevMaterials.3.033604 (2019).

144 Torrisi, S. B. *et al.* Random forest machine learning models for interpretable X-ray absorption near-edge structure spectrum-property relationships. *npj Computational Materials* **6**, doi:10.1038/s41524-020-00376-6 (2020).

145 Madkhali, M. M. M., Rankine, C. D. & Penfold, T. J. The Role of Structural Representation in the Performance of a Deep Neural Network for X-Ray Spectroscopy. *Molecules* **25**, doi:10.3390/molecules25112715 (2020).

146 Carbone, M. R., Topsakal, M., Lu, D. & Yoo, S. Machine-Learning X-Ray Absorption Spectra to Quantitative Accuracy. *Phys Rev Lett* **124**, 156401, doi:10.1103/PhysRevLett.124.156401 (2020).

147 Gilmer, J., Schoenholz, S. S., Riley, P. F., Vinyals, O. & Dahl, G. E. in *International Conference on Machine Learning.* 1263-1272 (PMLR).

148 Rankine, C. D., Madkhali, M. M. M. & Penfold, T. J. A Deep Neural Network for the Rapid Prediction of X-ray Absorption Spectra. *J Phys Chem A* **124**, 4263-4270, doi:10.1021/acs.jpca.0c03723 (2020).

149 Timoshenko, J., Lu, D., Lin, Y. & Frenkel, A. I. Supervised Machine-Learning-Based Determination of Three-Dimensional Structure of Metallic Nanoparticles. *J Phys Chem Lett* **8**, 5091-5098, doi:10.1021/acs.jpclett.7b02364 (2017).

150 Timoshenko, J. & Frenkel, A. I. "Inverting" X-ray Absorption Spectra of Catalysts by Machine Learning in Search for Activity Descriptors. *ACS Catalysis* **9**, 10192-10211, doi:10.1021/acscatal.9b03599 (2019).

151 Guda, A. A. *et al.* Quantitative structural determination of active sites from in situ and operando XANES spectra: From standard ab initio simulations to chemometric and machine learning approaches. *Catalysis Today* **336**, 3-21, doi:10.1016/j.cattod.2018.10.071 (2019).

152 Liu, Y. *et al.* Mapping XANES spectra on structural descriptors of copper oxide clusters using supervised machine learning. *J Chem Phys* **151**, 164201, doi:10.1063/1.5126597 (2019).

153 Marcella, N. *et al.* Neural network assisted analysis of bimetallic nanocatalysts using X-ray absorption near edge structure spectroscopy. *Phys Chem Chem Phys* **22**, 18902-18910, doi:10.1039/d0cp02098b (2020).
154 Trejo, O. *et al.* Elucidating the Evolving Atomic Structure in Atomic Layer Deposition Reactions with in Situ XANES and Machine Learning. *Chem Mater* **31**, 8937-8947, doi:10.1021/acs.chemmater.9b03025 (2019).
155 Suga, S. & Sekiyama, A. *Photoelectron spectroscopy : bulk and surface electronic structures*. (Heidelberg : Springer, [2013?], 2013).
156 Hofmann, S. *Auger- and X-ray photoelectron spectroscopy in materials science : a user-oriented guide*. (Springer, 2013).
157 Aarva, A., Deringer, V. L., Sainio, S., Laurila, T. & Caro, M. A. Understanding X-ray Spectroscopy of Carbonaceous Materials by Combining Experiments, Density Functional Theory, and Machine Learning. Part II: Quantitative Fitting of Spectra. *Chem Mater* **31**, 9256-9267, doi:10.1021/acs.chemmater.9b02050 (2019).
158 Drera, G., Kropf, C. M. & Sangaletti, L. Deep neural network for x-ray photoelectron spectroscopy data analysis. *Machine Learning: Science and Technology* **1**, 015008, doi:10.1088/2632-2153/ab5da6 (2020).
159 Medjanik, K. *et al.* Direct 3D mapping of the Fermi surface and Fermi velocity. *Nat Mater* **16**, 615-+, doi:10.1038/Nmat4875 (2017).
160 Moser, S. An experimentalist's guide to the matrix element in angle resolved photoemission. *J Electron Spectrosc* **214**, 29-52, doi:10.1016/j.elspec.2016.11.007 (2017).
161 He, Y., Wang, Y. & Shen, Z. X. Visualizing dispersive features in 2D image via minimum gradient method. *Review of Scientific Instruments* **88**, doi:Artn 073903,10.1063/1.4993919 (2017).
162 Peng, H. *et al.* Super resolution convolutional neural network for feature extraction in spectroscopic data. *Review of Scientific Instruments* **91**, doi:Artn 033905, 10.1063/1.5132586 (2020).
163 Xian, R. P. *et al.* A machine learning route between band mapping and band structure. *arXiv e-prints*, arXiv:2005.10210 (2020).
164 Furrer, A., Strassle, T. & Mesot, J. *Neutron scattering in condensed matter physics*. (World Scientific Pub., 2009).
165 Fernandez-Alonso, F. & Price, D. L. *Neutron scattering - Magnetic and Quantum Phenomena*. (Academic Press, an imprint of Elsevier, 2015).
166 Squires, G. L. *Introduction to the theory of thermal neutron scattering*. (Cambridge University Press, 1978).
167 Shirane, G., Shapiro, S. M. & Tranquada, J. M. *Neutron scattering with a triple-axis spectrometer : basic techniques*. (Cambridge University Press, 2002).
168 Dorner, B. *Coherent inelastic neutron scattering in lattice dynamics*. (Springer-Verlag, 1982).
169 Sinn, H. Spectroscopy with meV energy resolution. *J Phys-Condens Mat* **13**, 7525-7537, doi:Doi 10.1088/0953-8984/13/34/305 (2001).
170 Burkel, E. Determination of phonon dispersion curves by means of inelastic x-ray scattering. *J Phys-Condens Mat* **13**, 7627-7644, doi:Doi 10.1088/0953-8984/13/34/310 (2001).

171 Nguyen, T. *et al.* Topological Singularity Induced Chiral Kohn Anomaly in a Weyl Semimetal. *Phys Rev Lett* **124**, 236401, doi:10.1103/PhysRevLett.124.236401 (2020).
172 Born, M. & Huang, K. *Dynamical theory of crystal lattices*. (Clarendon Press; Oxford University Press, 1988).
173 Taraskin, S. N. & Elliott, S. R. Phonons in vitreous silica: Dispersion and localization. *Europhys Lett* **39**, 37-42 (1997).
174 Novikov, V. N. & Surovtsev, N. V. Spatial structure of boson peak vibrations in glasses. *Physical Review B* **59**, 38-41, doi:DOI 10.1103/PhysRevB.59.38 (1999).
175 Chumakov, A. I. *et al.* Collective nature of the boson peak and universal transboson dynamics of glasses. *Phys Rev Lett* **92**, 245508, doi:10.1103/PhysRevLett.92.245508 (2004).
176 Shintani, H. & Tanaka, H. Universal link between the boson peak and transverse phonons in glass. *Nat Mater* **7**, 870-877, doi:10.1038/nmat2293 (2008).
177 Chumakov, A. I. *et al.* Equivalence of the boson peak in glasses to the transverse acoustic van Hove singularity in crystals. *Phys Rev Lett* **106**, 225501, doi:10.1103/PhysRevLett.106.225501 (2011).
178 Sokoloff, J. B. Theory of Inelastic Neutron Scattering in the Itinerant Model Antiferromagnetic Metals. I. *Phys Rev* **185**, 770-782, doi:10.1103/PhysRev.185.770 (1969).
179 Lee, S. H. *et al.* Emergent excitations in a geometrically frustrated magnet. *Nature* **418**, 856-858, doi:10.1038/nature00964 (2002).
180 Helton, J. S. *et al.* Spin dynamics of the spin-1/2 kagome lattice antiferromagnet ZnCu3(OH)6Cl2. *Phys Rev Lett* **98**, 107204, doi:10.1103/PhysRevLett.98.107204 (2007).
181 Banerjee, A. *et al.* Neutron scattering in the proximate quantum spin liquid alpha-RuCl3. *Science* **356**, 1055-1058 (2017).
182 Eschrig, M. The effect of collective spin-1 excitations on electronic spectra in high-T-c superconductors. *Adv Phys* **55**, 47-183, doi:Doi 10.1080/00018730600645636 (2006).
183 Christianson, A. D. *et al.* Unconventional superconductivity in Ba(0.6)K(0.4)Fe2As2 from inelastic neutron scattering. *Nature* **456**, 930-932, doi:10.1038/nature07625 (2008).
184 Li, Y. *et al.* Hidden magnetic excitation in the pseudogap phase of a high-T(c) superconductor. *Nature* **468**, 283-285, doi:10.1038/nature09477 (2010).
185 Chen, Z. *et al.* Direct prediction of phonon density of states with Euclidean neural network. *arXiv e-prints*, arXiv:2009.05163 (2020).
186 Petretto, G. *et al.* High-throughput density-functional perturbation theory phonons for inorganic materials. *Scientific Data* **5**, 180065, doi:10.1038/sdata.2018.65 (2018).
187 Smidt, T. E., Geiger, M. & Miller, B. K. Finding Symmetry Breaking Order Parameters with Euclidean Neural Networks. *arXiv e-prints*, arXiv:2007.02005 (2020).
188 Toth, S. & Lake, B. Linear spin wave theory for single-Q incommensurate magnetic structures. *Journal of physics. Condensed matter : an Institute of Physics journal* **27**, 166002, doi:10.1088/0953-8984/27/16/166002 (2015).

189 Li, Y., Cheng, W., Yu, L. H. & Rainer, R. Genetic algorithm enhanced by machine learning in dynamic aperture optimization. *Physical Review Accelerators and Beams* **21**, doi:10.1103/PhysRevAccelBeams.21.054601 (2018).

190 Leemann, S. C. *et al.* Demonstration of Machine Learning-Based Model-Independent Stabilization of Source Properties in Synchrotron Light Sources. *Phys Rev Lett* **123**, 194801, doi:10.1103/PhysRevLett.123.194801 (2019).

191 Wan, J., Chu, P., Jiao, Y. & Li, Y. Improvement of machine learning enhanced genetic algorithm for nonlinear beam dynamics optimization. *Nuclear Instruments and Methods in Physics Research Section A: Accelerators, Spectrometers, Detectors and Associated Equipment* **946**, 162683, doi:10.1016/j.nima.2019.162683 (2019).

192 Piekarski, M., Jaworek-Korjakowska, J., Wawrzyniak, A. I. & Gorgon, M. Convolutional neural network architecture for beam instabilities identification in Synchrotron Radiation Systems as an anomaly detection problem. *Measurement* **165**, 108116, doi:10.1016/j.measurement.2020.108116 (2020).

193 Asahara, A. *et al.* Early-Stopping of Scattering Pattern Observation with Bayesian Modeling. *Thirty-Third Aaai Conference on Artificial Intelligence / Thirty-First Innovative Applications of Artificial Intelligence Conference / Ninth Aaai Symposium on Educational Advances in Artificial Intelligence,* 9410-9415 (2019).

194 Kanazawa, T., Asahara, A. & Morita, H. Accelerating small-angle scattering experiments with simulation-based machine learning. *Journal of Physics: Materials* **3**, 015001, doi:10.1088/2515-7639/ab3c45 (2019).

195 Noack, M. M. *et al.* A Kriging-Based Approach to Autonomous Experimentation with Applications to X-Ray Scattering. *Sci Rep* **9**, 11809, doi:10.1038/s41598-019-48114-3 (2019).

196 Noack, M. M., Doerk, G. S., Li, R., Fukuto, M. & Yager, K. G. Advances in Kriging-Based Autonomous X-Ray Scattering Experiments. *Sci Rep* **10**, 1325, doi:10.1038/s41598-020-57887-x (2020).

197 Cressie, N. The origins of kriging. *Mathematical geology* **32**, 239--252 (1990).

198 Rasmussen, C. E. & Williams, C. K. I. *Gaussian Processes for Machine Learning*. (MIT Press, 2006).

199 Yang, X., De Carlo, F., Phatak, C. & Gursoy, D. A convolutional neural network approach to calibrating the rotation axis for X-ray computed tomography. *J Synchrotron Radiat* **24**, 469-475, doi:10.1107/S1600577516020117 (2017).

200 Lee, S. *et al.* Deep learning for high-resolution and high-sensitivity interferometric phase contrast imaging. *Sci Rep* **10**, 9891, doi:10.1038/s41598-020-66690-7 (2020).

201 Shashank Kaira, C. *et al.* Automated correlative segmentation of large Transmission X-ray Microscopy (TXM) tomograms using deep learning. *Materials Characterization* **142**, 203-210, doi:10.1016/j.matchar.2018.05.053 (2018).

202 Hui, Y. & Liu, Y. Volumetric Data Exploration with Machine Learning-Aided Visualization in Neutron Science. *Advances in Computer Vision* **943**, 257-271, doi:10.1007/978-3-030-17795-9_18 (2020).

203 Yang, X. *et al.* Low-dose x-ray tomography through a deep convolutional neural network. *Sci Rep* **8**, 2575, doi:10.1038/s41598-018-19426-7 (2018).

204 Zhang, R., Liu, Y. & Sun, H. Physics-informed multi-LSTM networks for metamodeling of nonlinear structures. *Computer Methods in Applied Mechanics and Engineering* **369**, doi:10.1016/j.cma.2020.113226 (2020).

205 Flurin, E., Martin, L. S., Hacohen-Gourgy, S. & Siddiqi, I. Using a Recurrent Neural Network to Reconstruct Quantum Dynamics of a Superconducting Qubit from Physical Observations. *Physical Review X* **10**, doi:10.1103/PhysRevX.10.011006 (2020).

206 Roth, C. Iterative retraining of quantum spin models using recurrent neural networks. *arXiv preprint arXiv:2003.06228* (2020).

207 Deng, A., Huang, J., Liu, H. & Cai, W. Deep learning algorithms for temperature field reconstruction of nonlinear tomographic absorption spectroscopy. *Measurement: Sensors* **10-12**, doi:10.1016/j.measen.2020.100024 (2020).

208 Simine, L., Allen, T. C. & Rossky, P. J. Predicting optical spectra for optoelectronic polymers using coarse-grained models and recurrent neural networks. *Proc Natl Acad Sci U S A* **117**, 13945-13948, doi:10.1073/pnas.1918696117 (2020).

209 Tian, C. *et al.* Deep learning on image denoising: An overview. *Neural Netw* **131**, 251-275, doi:10.1016/j.neunet.2020.07.025 (2020).

210 Deng, M., Li, S., Goy, A., Kang, I. & Barbastathis, G. Learning to synthesize: robust phase retrieval at low photon counts. *Light Sci Appl* **9**, 36, doi:10.1038/s41377-020-0267-2 (2020).

211 Goy, A., Arthur, K., Li, S. & Barbastathis, G. Low Photon Count Phase Retrieval Using Deep Learning. *Phys Rev Lett* **121**, 243902, doi:10.1103/PhysRevLett.121.243902 (2018).

212 Che, Z., Purushotham, S., Cho, K., Sontag, D. & Liu, Y. Recurrent Neural Networks for Multivariate Time Series with Missing Values. *Sci Rep* **8**, 6085, doi:10.1038/s41598-018-24271-9 (2018).

213 Chen, R. T., Rubanova, Y., Bettencourt, J. & Duvenaud, D. Neural ordinary differential equations. *arXiv preprint arXiv:1806.07366* (2018).

214 Hashimoto, K., Hu, H.-Y. & You, Y.-Z. Neural ODE and holographic QCD. *arXiv preprint arXiv:2006.00712* (2020).

215 Nakajima, M., Tanaka, K. & Hashimoto, T. Neural Schr$\backslash$"$\{$o$\}$ dinger Equation: Physical Law as Neural Network. *arXiv preprint arXiv:2006.13541* (2020).

216 Lusch, B., Kutz, J. N. & Brunton, S. L. Deep learning for universal linear embeddings of nonlinear dynamics. *Nature Communications* **9**, doi:10.1038/s41467-018-07210-0 (2018).

217 Champion, K., Lusch, B., Kutz, J. N. & Brunton, S. L. Data-driven discovery of coordinates and governing equations. *Proc Natl Acad Sci U S A* **116**, 22445-22451, doi:10.1073/pnas.1906995116 (2019).

218 Ngiam, J. *et al.* in *ICML*.

219 Sohn, K., Shang, W. & Lee, H. Improved multimodal deep learning with variation of information. *Advances in neural information processing systems* **27**, 2141-2149 (2014).

220 Baltrušaitis, T., Ahuja, C. & Morency, L.-P. Multimodal machine learning: A survey and taxonomy. *IEEE transactions on pattern analysis and machine intelligence* **41**, 423-443 (2018).

221 Liu, K., Li, Y., Xu, N. & Natarajan, P. Learn to combine modalities in multimodal deep learning. *arXiv preprint arXiv:1805.11730* (2018).

222 Radu, V. *et al.* in *Proceedings of the 2016 ACM International Joint Conference on Pervasive and Ubiquitous Computing: Adjunct* 185–188 (Association for Computing Machinery, Heidelberg, Germany, 2016).

223 Radu, V. *et al.* Multimodal Deep Learning for Activity and Context Recognition. *Proc. ACM Interact. Mob. Wearable Ubiquitous Technol.* **1**, Article 157, doi:10.1145/3161174 (2018).

224 Eitel, A., Springenberg, J. T., Spinello, L., Riedmiller, M. & Burgard, W. in *2015 IEEE/RSJ International Conference on Intelligent Robots and Systems (IROS).* 681-687.

225 Hong, D. *et al.* More Diverse Means Better: Multimodal Deep Learning Meets Remote-Sensing Imagery Classification. *IEEE Transactions on Geoscience and Remote Sensing*, 1-15, doi:10.1109/TGRS.2020.3016820 (2020).

226 Liu, W., Zheng, W.-L. & Lu, B.-L. in *International conference on neural information processing.* 521-529 (Springer).

227 Liang, M., Li, Z., Chen, T. & Zeng, J. Integrative data analysis of multi-platform cancer data with a multimodal deep learning approach. *IEEE/ACM transactions on computational biology and bioinformatics* **12**, 928-937 (2014).

228 Xu, T., Zhang, H., Huang, X., Zhang, S. & Metaxas, D. N. in *International conference on medical image computing and computer-assisted intervention.* 115-123 (Springer).

229 Hybertsen, M. S. & Louie, S. G. First-Principles Theory of Quasiparticles: Calculation of Band Gaps in Semiconductors and Insulators. *Phys Rev Lett* **55**, 1418-1421, doi:10.1103/physrevlett.55.1418 (1985).

230 Anisimov, V. I., Poteryaev, A. I., Korotin, M. A., Anokhin, A. O. & Kotliar, G. First-principles calculations of the electronic structure and spectra of strongly correlated systems: dynamical mean-field theory. *Journal of Physics: Condensed Matter* **9**, 7359-7367, doi:10.1088/0953-8984/9/35/010 (1997).

231 Georges, A. & Kotliar, G. Hubbard model in infinite dimensions. *Physical Review B* **45**, 6479-6483, doi:10.1103/physrevb.45.6479 (1992).

232 Haule, K. & Kotliar, G. Coherence–incoherence crossover in the normal state of iron oxypnictides and importance of Hund's rule coupling. *New J Phys* **11**, doi:10.1088/1367-2630/11/2/025021 (2009).

233 Park, H., Haule, K. & Kotliar, G. Magnetic Excitation Spectra in BaFe2As2: A Two-Particle Approach within a Combination of the Density Functional Theory and the Dynamical Mean-Field Theory Method. *Phys Rev Lett* **107**, doi:10.1103/physrevlett.107.137007 (2011).

234 Hariki, A., Winder, M., Uozumi, T. & Kuneš, J. LDA+DMFT approach to resonant inelastic x-ray scattering in correlated materials. *Physical Review B* **101**, doi:10.1103/physrevb.101.115130 (2020).

235 Li, Q., Callaway, J. & Tan, L. Spectral function in the two-dimensional Hubbard model. *Physical Review B* **44**, 10256-10269, doi:10.1103/physrevb.44.10256 (1991).
236 Bartlett, R. J. & Musiał, M. Coupled-cluster theory in quantum chemistry. *Rev Mod Phys* **79**, 291-352, doi:10.1103/revmodphys.79.291 (2007).
237 White, S. R. & Feiguin, A. E. Real-Time Evolution Using the Density Matrix Renormalization Group. *Phys Rev Lett* **93**, doi:10.1103/physrevlett.93.076401 (2004).
238 Foulkes, W. M. C., Mitas, L., Needs, R. J. & Rajagopal, G. Quantum Monte Carlo simulations of solids. *Rev Mod Phys* **73**, 33-83, doi:10.1103/revmodphys.73.33 (2001).
239 Winter, S. M. *et al.* Breakdown of magnons in a strongly spin-orbital coupled magnet. *Nature Communications* **8**, doi:10.1038/s41467-017-01177-0 (2017).
240 Huang, E. W. *et al.* Numerical evidence of fluctuating stripes in the normal state of high-Tccuprate superconductors. *Science* **358**, 1161-1164, doi:10.1126/science.aak9546 (2017).
241 Carleo, G. & Troyer, M. Solving the quantum many-body problem with artificial neural networks. *Science* **355**, 602-606, doi:10.1126/science.aag2302 (2017).
242 Von Lilienfeld, O. A., Müller, K.-R. & Tkatchenko, A. Exploring chemical compound space with quantum-based machine learning. *Nature Reviews Chemistry* **4**, 347-358, doi:10.1038/s41570-020-0189-9 (2020).

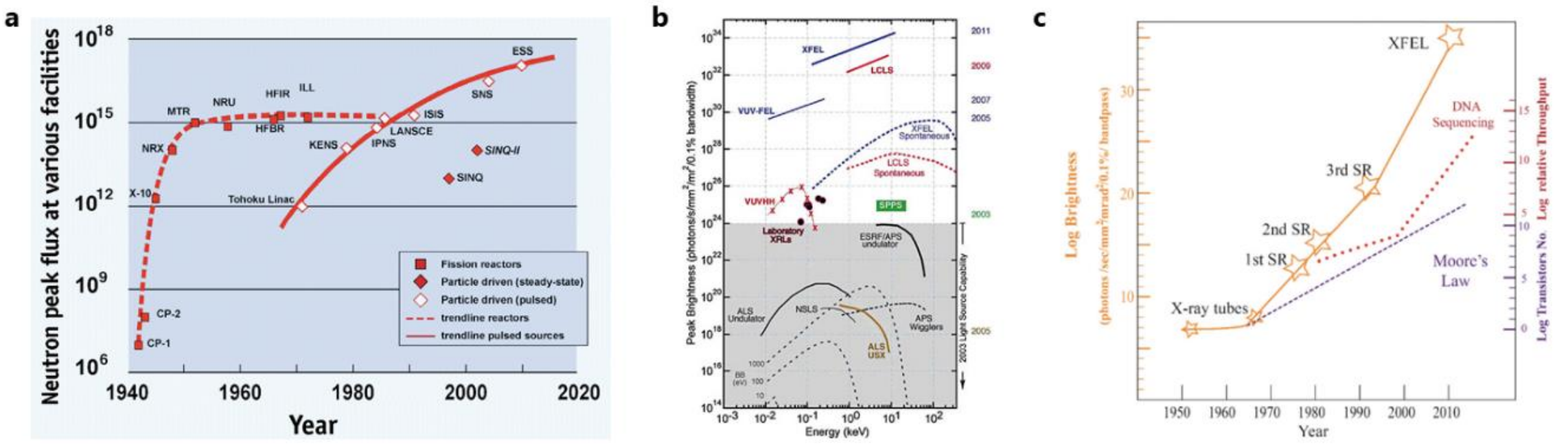

**Figure 1. Neutron and X-ray scattering in the data era. a.** Improvement of neutron flux in reactor-based neutron generation (dashed-line) and accelerator-based spallation neutron generation (solid-line). Figure reproduced from Bohn *et al.*[3]. **b.** Drastic enhancement of peak brightness in X-ray scattering at various energy scales. Figure reproduced from Ref[4]. **c.** Comparison of increasing speed of peak brightness of synchrotron X-ray scattering and the Moore's law of microelectronics. Figure reproduced from Su *et al.*[6].

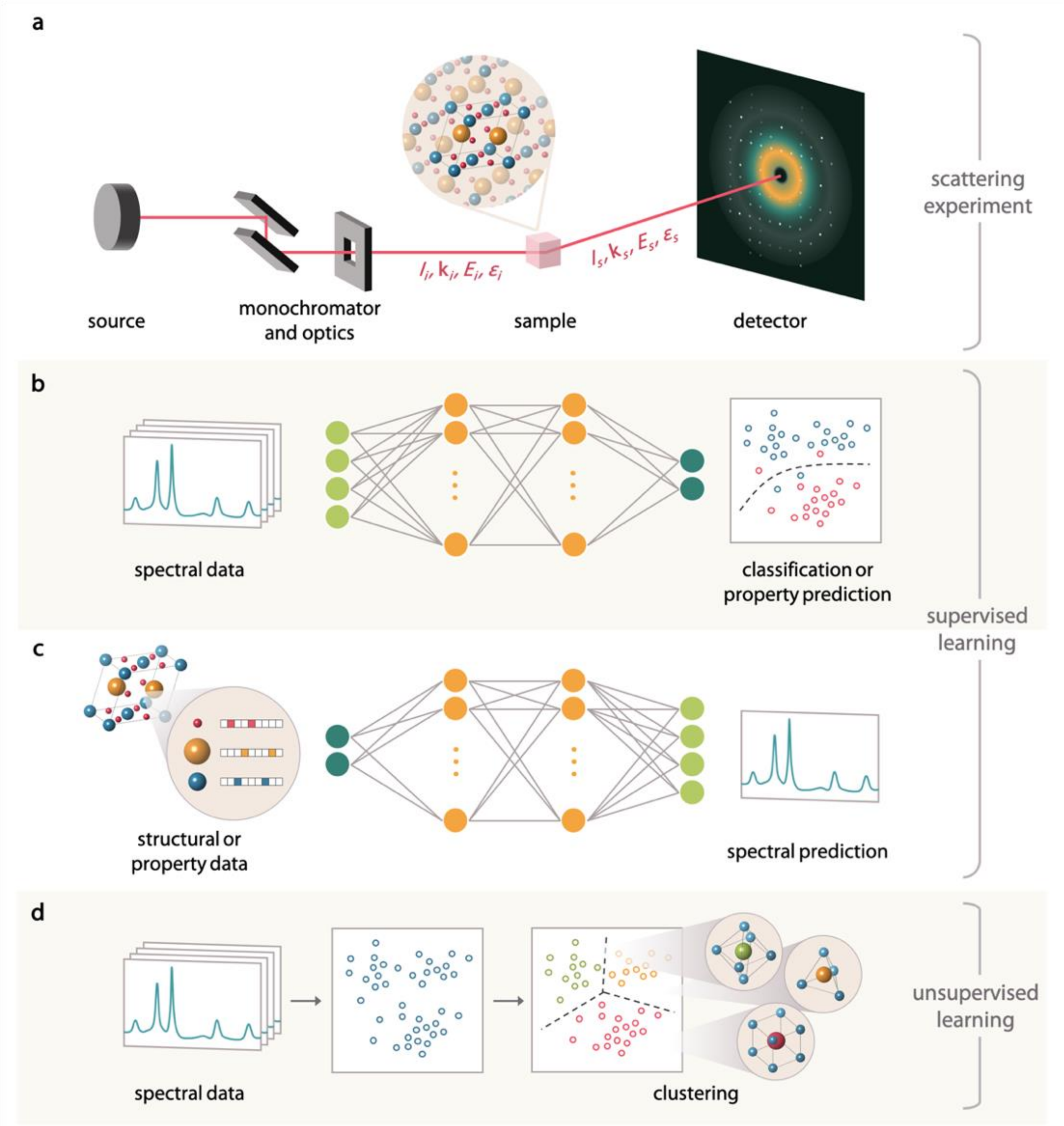

**Figure 2. Machine learning in a neutron and X-ray scattering pipeline. a.** Schematic of a typical scattering setup **b.** Spectral data serving as input to a supervised machine learning model for a materials' classification or property prediction task. **c.** Materials structural or property data serving as input to a supervised machine learning model for direct and efficient prediction of the full scattering spectra. **d.** Spectral data as part of an unsupervised machine learning task that can identify inherent patterns or clusters within the dataset that may correspond to meaningful physical parameters.

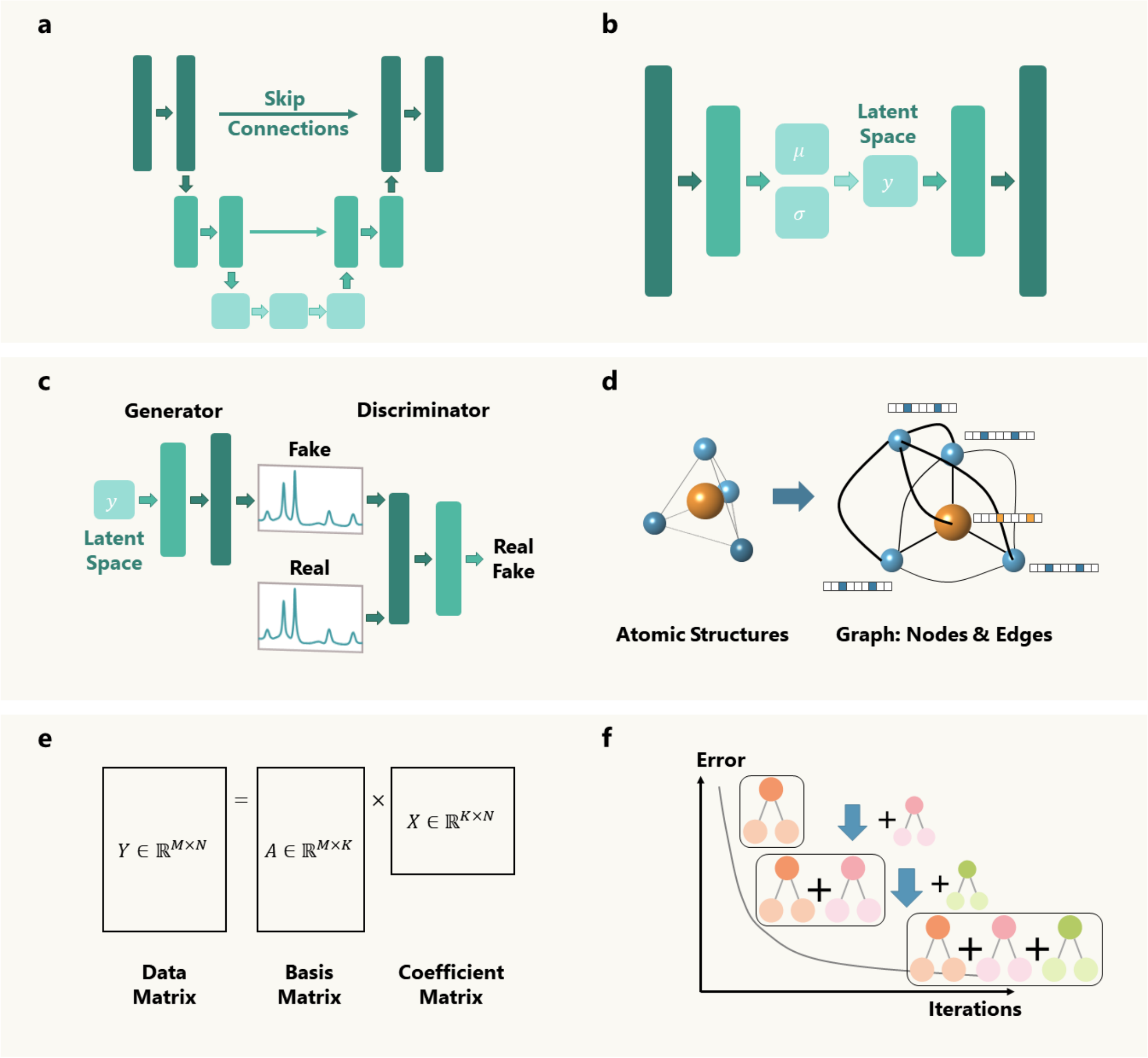

**Figure 3. Common machine learning architectures. a.** U-Net architecture. **b.** Variational autoencoder (VAE). **c.** Generative adversarial network (GAN). **d.** Crystal graph convolutional neural network (CGCNN). **e.** Non-negative matrix factorization (NMF). **f.** Gradient boosting trees (GB Trees).

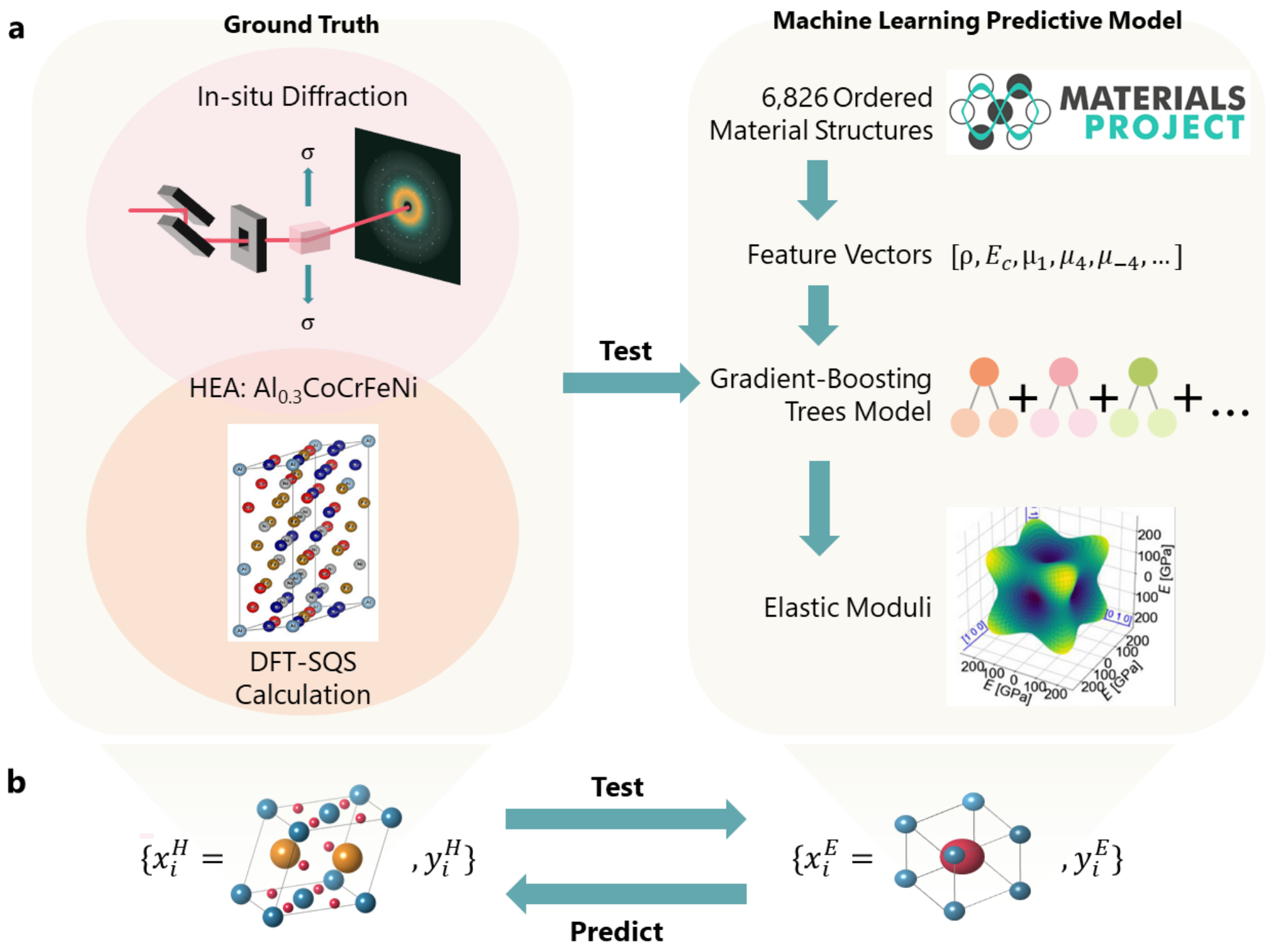

**Figure 4. Machine learning models for predicting properties that are hard to acquire. a.** Predict elastic moduli of high entropy alloys (HEA) using the machine learning based model that is trained with ordered crystalline solids. **b.** Machine learning models can be trained with data that are from or augmented by easily accessible samples, then the models can be used to predict properties that are hard to acquire. Figure adapted from Kim *et al*.[74].

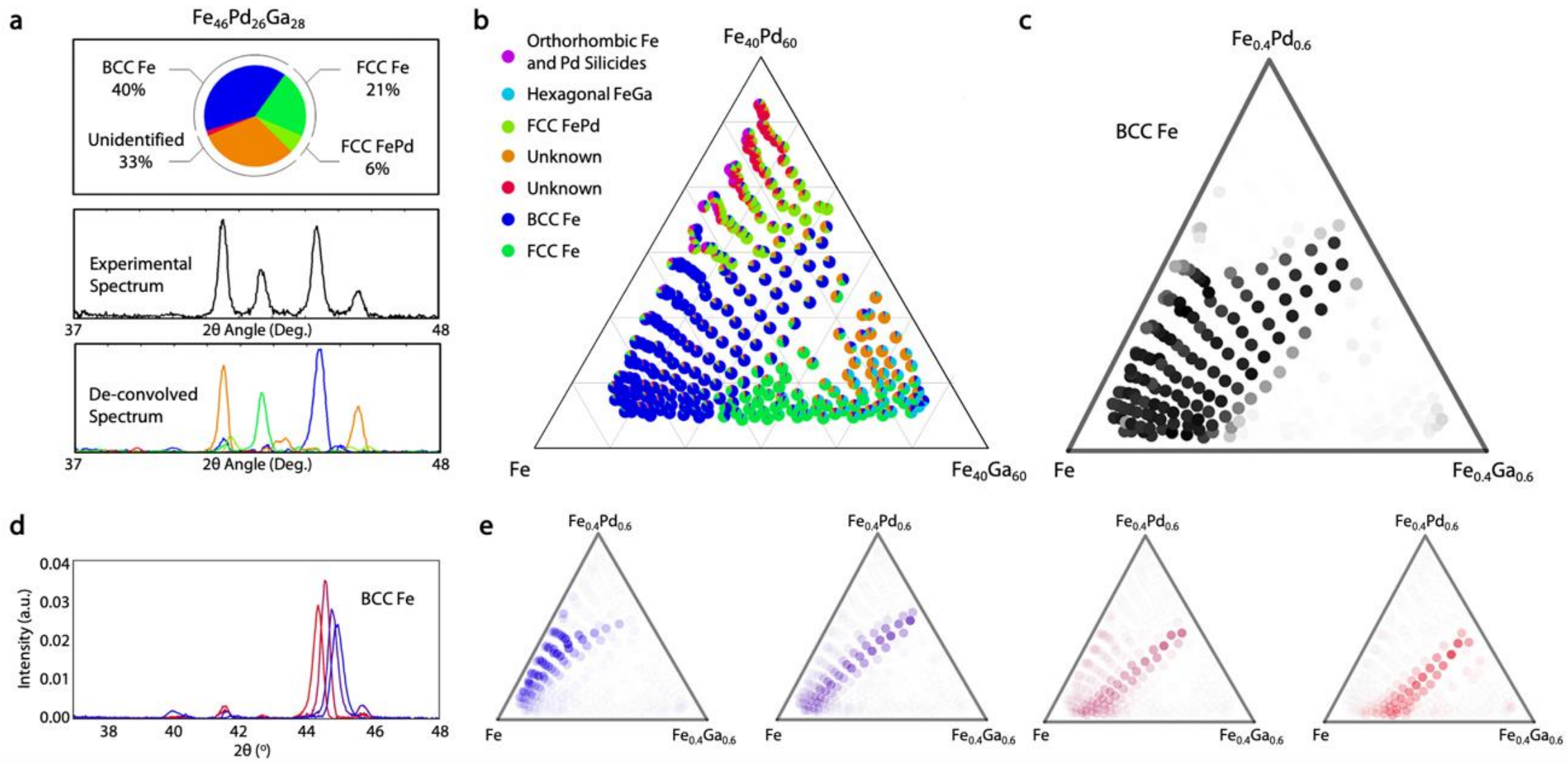

**Figure 5. Structural phase mapping of X-ray diffraction data with nonnegative matrix factorization (NMF). (a-b). a.** Weights of the basis patterns for $Fe_{46}Pd_{26}Ga_{28}$ (top panel), and decomposition of the experimental XRD spectrum (middle panel) into the weighted basis patterns (bottom panel). **b.** Structural phase diagram of the Fe-Ga-Pd system constructed using the weights of the basis diffraction patterns found by NMF. Subfigures **a.** and **b.** are reproduced from Long *et al.*[79]. **(c-e). c.** Structural phase diagram of the Fe-Ga-Pd system reconstructed using NMF with custom clustering (NMFk). **d.** Four basis patterns of the Fe-Ga-Pd system obtained by NMFk representing the BCC Fe phase, which differ only in peak-shift. **e.** The weight of each basis pattern from **d.** drawn in the corresponding color, showing the peak evolution in the composition phase space. Subfigures **c-e** are reproduced from Stanev *et* al.[80].

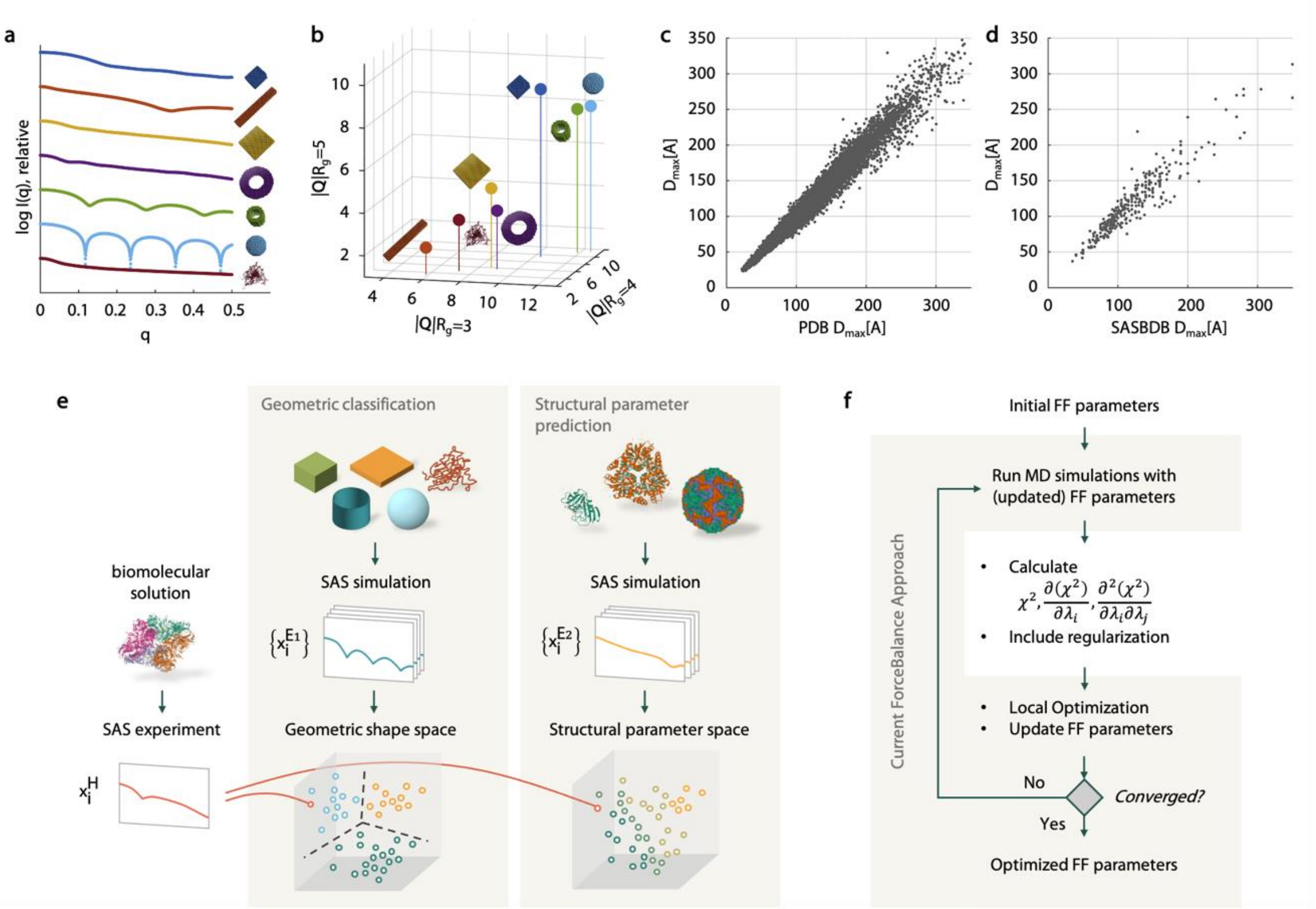

**Figure 6. Machine learning for small-angle scattering. (a-d).** Reproduced from Franke *et al.*[102]. **a.** Small-angle scattering patterns of geometric objects and a disordered chain. **b.** Reduction of each scattering pattern in **a** to three features, corresponding to each axis of the three-dimensional space, and an associated class label. **(c-d)**. Estimates of the structural parameter $D_{max}$ for the **c.** Protein Data Bank (PDB) dataset and **d.** experimental SAS dataset from SASBDB, compared to their expected values. **e.** Schematic illustration of the workflow followed by Franke *et al.* A model for geometric classification is obtained by reducing the simulated SAS patterns of geometric objects and disordered chains to three features to construct a geometric shape space, with each datapoint labeled according to its associated geometric class. Similarly, a model for structural parameter prediction is obtained by reducing the simulated SAS patterns of asymmetric units and biological assemblies available in the PDB to the same three features. Each datapoint in this structural parameter space is associated with a value for each structural parameter of interest, such as maximal extent ($D_{max}$) and molecular mass. Subsequently, the geometry and structural parameters of an entity in an unknown biomolecular solution can be determined by computing the three features using its experimental SAS spectrum, mapping to the corresponding coordinate in either geometric or structural parameter spaces, and weighting the contributions of its *k*-nearest neighbors. **f.** Flowchart depicting the modified ForceBalance-SAS algorithm. Adapted from Demerdash *et al.*[103].

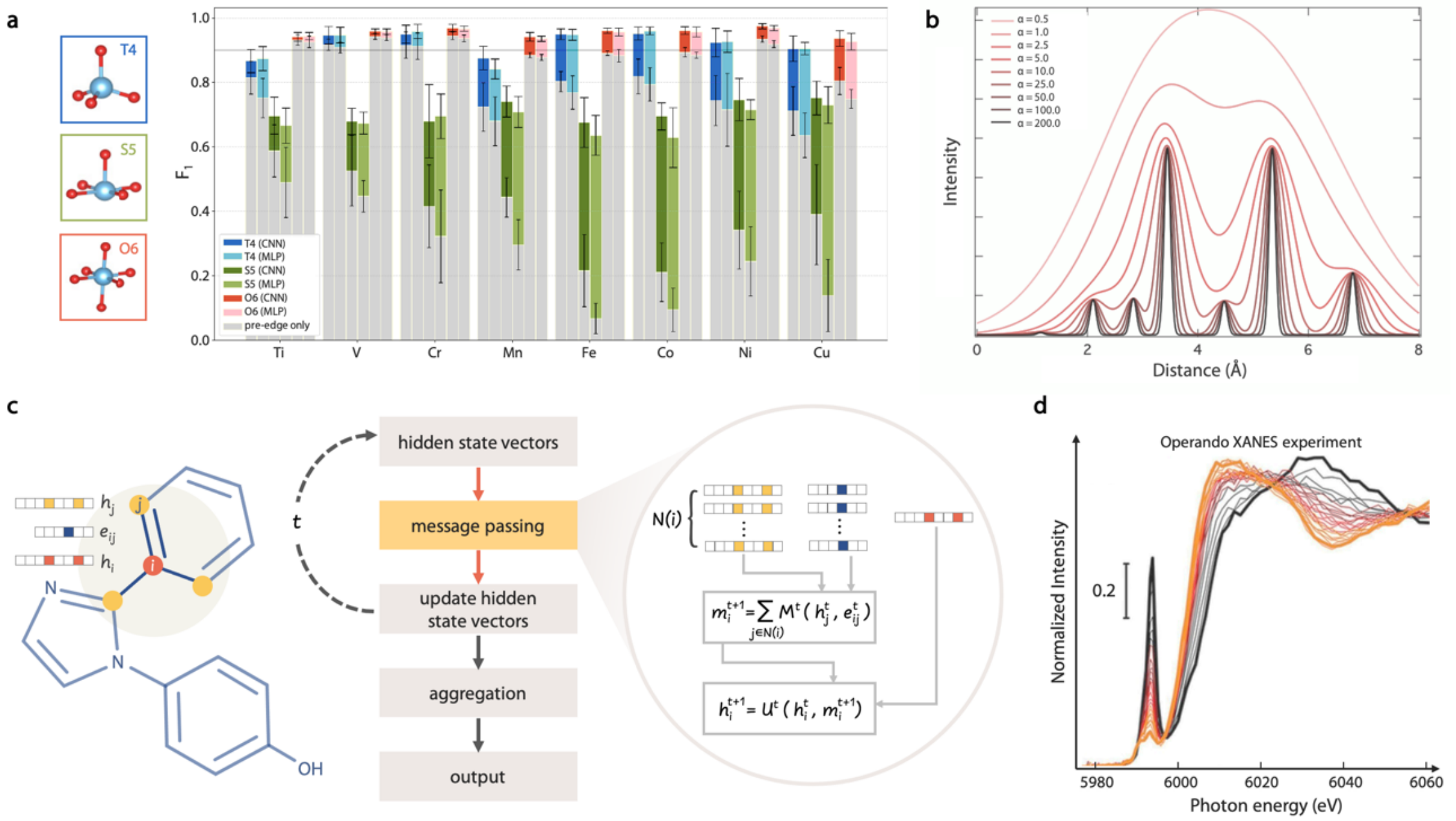

**Figure 7. Machine learning for X-ray absorption spectroscopy. a.** $F_1$ scores for the classification accuracy for each coordination environment – tetrahedral (T4), square pyramidal (S5), and octahedral (O6) – depicted at left, using two different machine learning models (CNN and MLP). The full bar height represents the score for models trained on the full XANES feature space, while the gray bars represent the results using only the pre-edge region. Adapted from Carbone *et al.*[143]. **b.** Representative RDC for an arbitrary system for nine different values of α, depicting the increasing resolution of the RDC with increasing α. Adapted from Madkhali *et al.*[145]. **c.** Schematic illustration of a message-passing neural network (MPNN). **d.** Time evolution of the *operando* Cr K-edge XANES spectra of a $Cr^{VI}/SiO_2$ catalyst during reduction with ethylene (from black to red) and during ethylene polymerization (from red to bold orange). Adapted from Guda *et al.*[151].

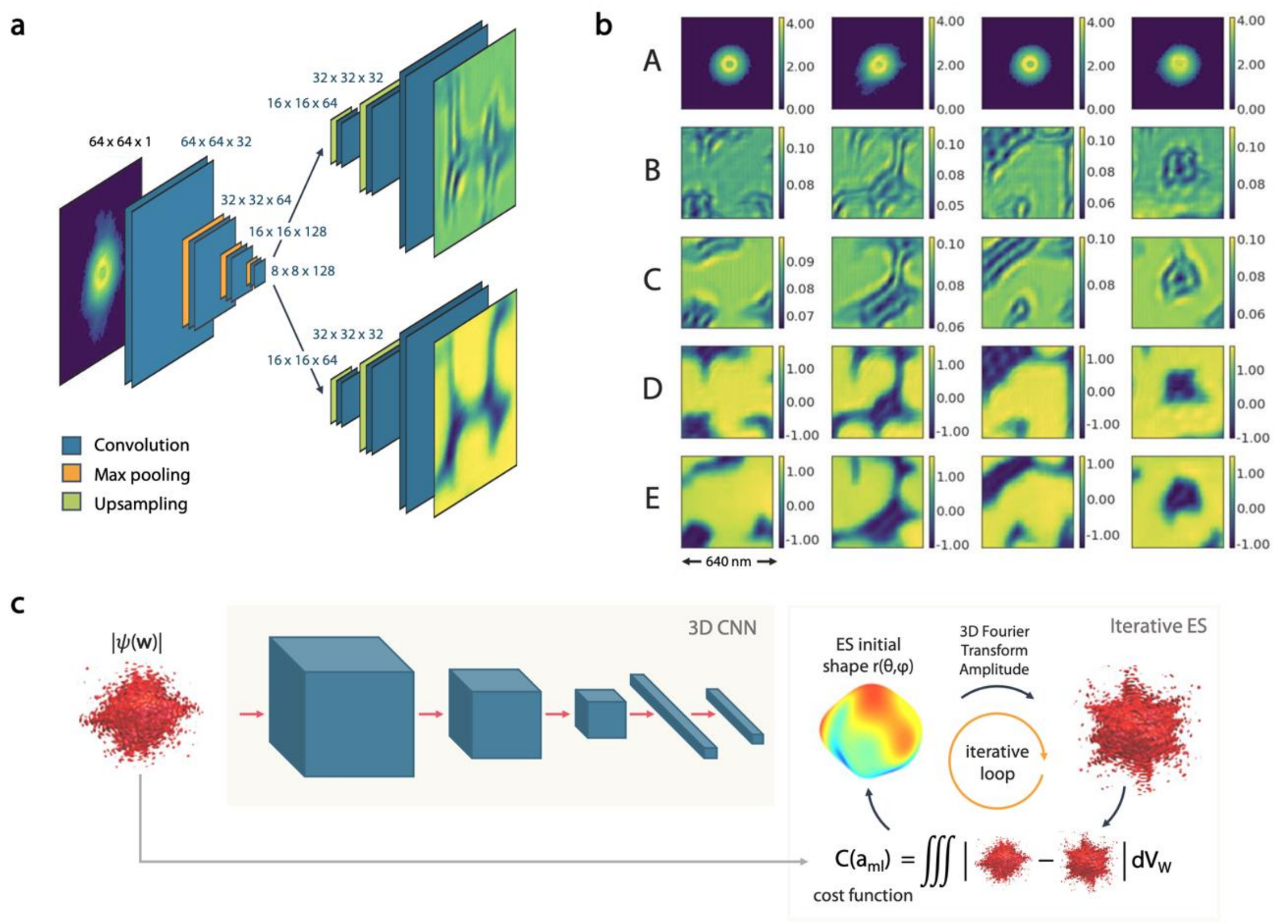

**Figure 8. Machine learning for coherent diffraction imaging. a.** Architecture of PtychoNN, a CNN-based model that directly solves the phase retrieval problem in ptychography. Adapted from Cherukara *et al.*[134]. **b.** The PtychoNN takes diffraction patterns at different spatial scan points (row A), and retrieve real space amplitude and phase (row C and E), which show good agreement with conventional phase retrieval algorithm approach (row B and D). Adapted from Cherukara *et al.*[134]. **c.** An adaptive machine learning framework that recovers real space electron densities from diffraction patterns. This framework takes output of the 3D CNN as initial condition of the extremum seeking algorithm. Adapted from Scheinker and Pokharel[135].

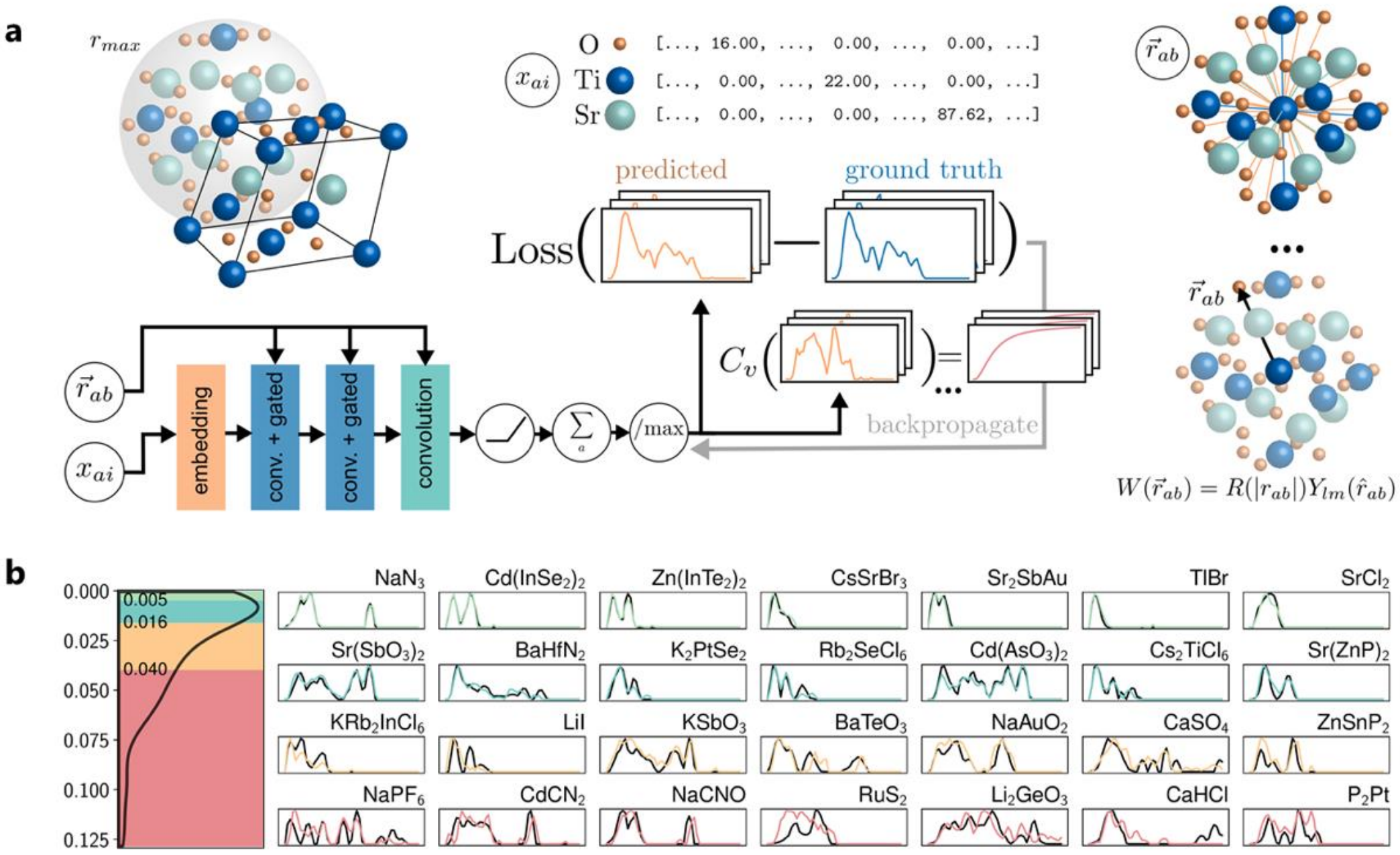

**Figure 9. Direct prediction of phonon density of states (DOS) with the Euclidean neural network ($E^3$NN). a.** Crystal structures is encoded into graphs with atoms being nodes with feature vectors, and bonds being edges. The graph is then passed into the $E^3$NN that preserves the crystallographic symmetry. **b.** The predicted phonon DOS is displayed in four rows with each representing an error quartile. Fine details of DOS can be well captured for samples in first three rows (lower error predictions), while coarse features such as bandwidth and DOS gap can still be largely predicted for last row (higher error predictions). Reproduced from Chen, Andrejevic, and Smidt *et al.*[185].

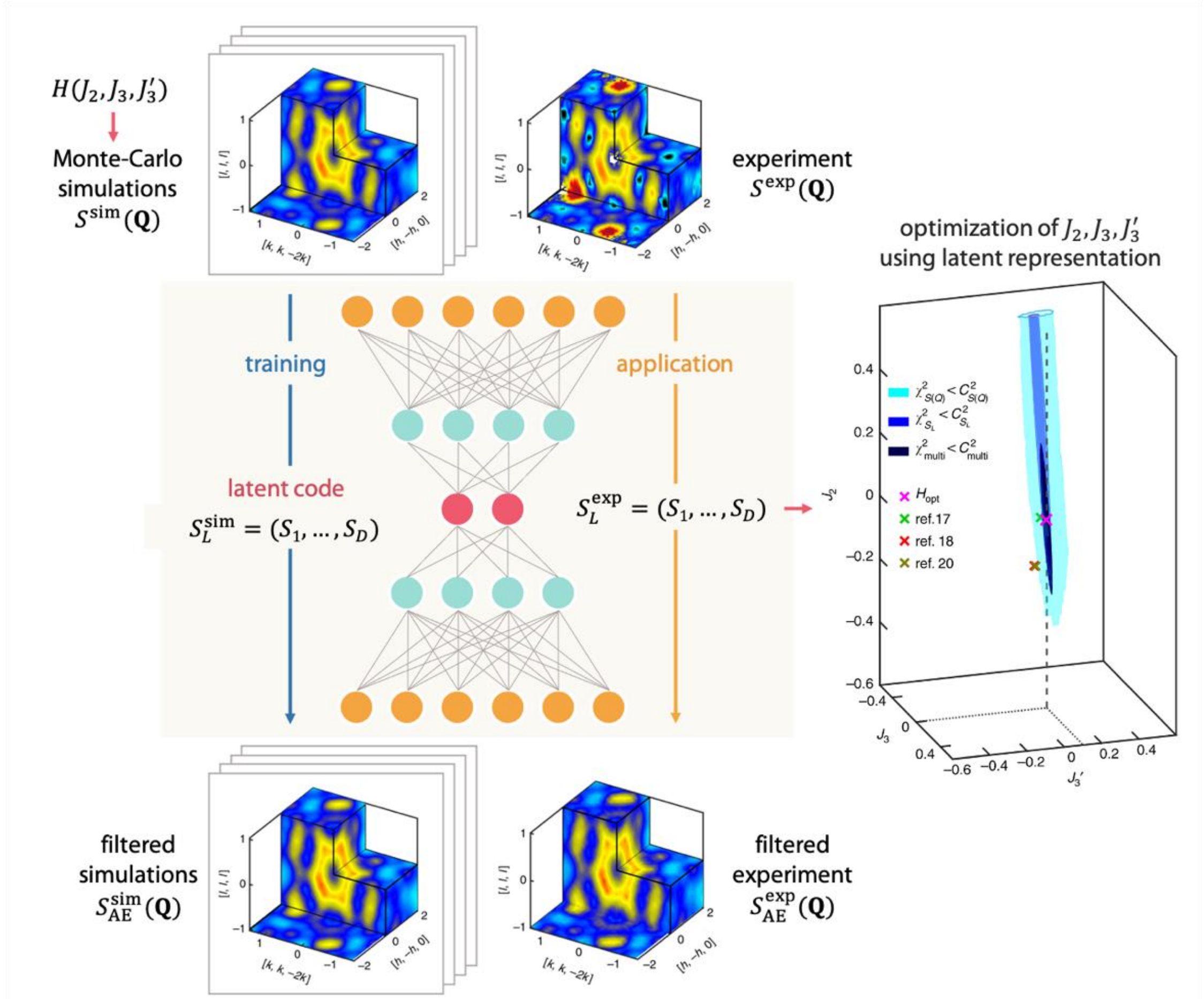

**Figure 10. Autoencoder-assisted Hamiltonian parameter estimation for experimentally measured magnetic structure factor $S^{\text{exp}}(\mathbf{Q})$.** The autoencoder is trained with 1,000 Monte-Carlo simulated data $S^{\text{sim}}(\mathbf{Q})$ from the spin-ice Hamiltonian $H(J_2, J_3, J_3')$. To find a set of parameters $\{J_2, J_3, J_3'\}$ that best represent experimental data, a cost function containing squared Euclidean distance in the learned latent space, namely $\left\| S_L^{\text{exp}} - S_L^{\text{sim}} \right\|^2$, is minimized. Reproduced from Samarakoon *et al.*[67].

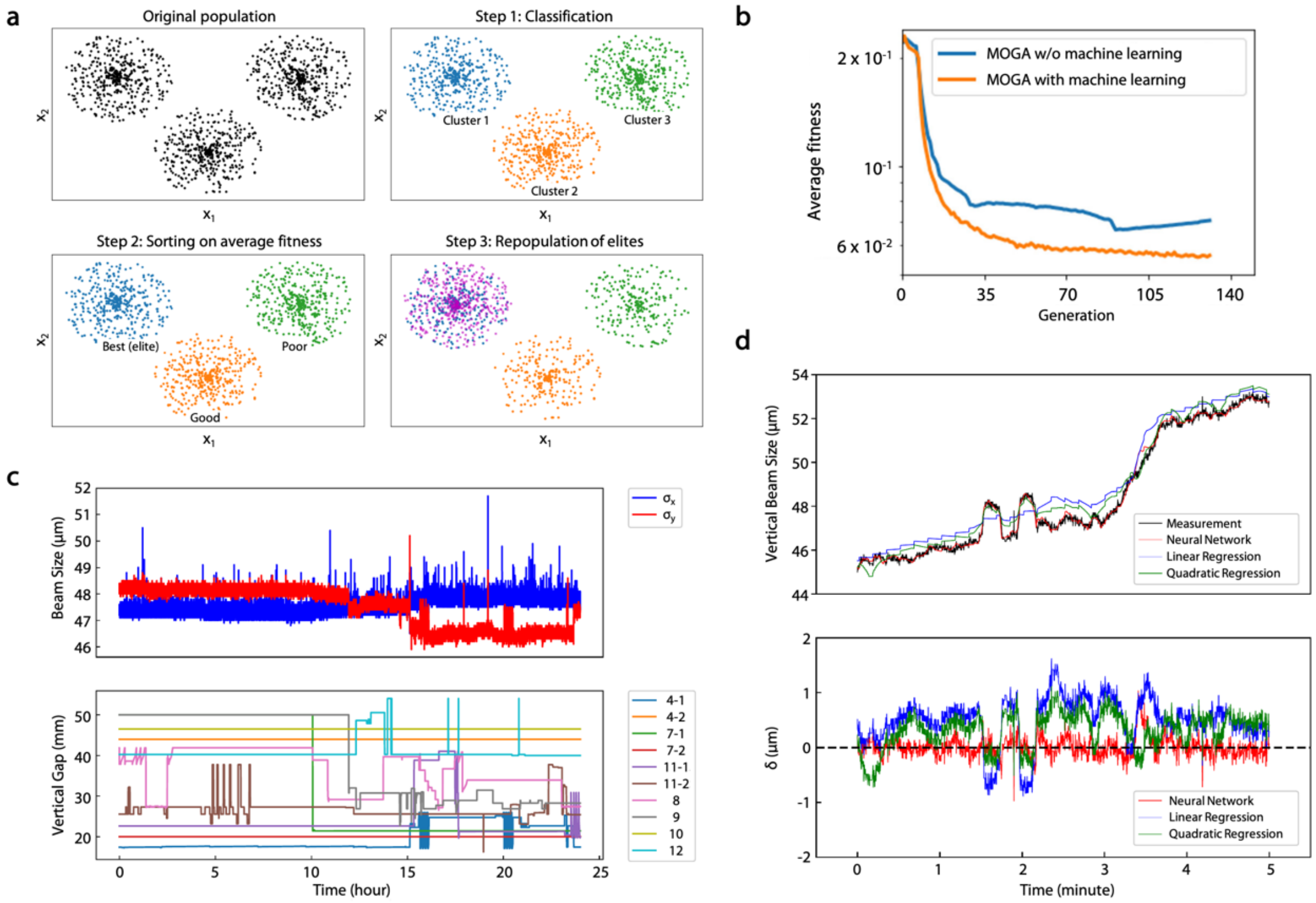

**Figure 11. Machine learning applications in scattering instruments and beams. a.** *K*-means clustering method is applied to classify parameter populations, where the regime of desired parameters is used to help to generate new candidates. **b.** Faster convergence towards optimized parameters can be achieved with machine learning. Subfigures **a.** and **b.** are reproduced from Li *et al.*[189]. **c.** Variation in electron beam size (top) exists due to insertion device gaps (bottom). **d.** A neural network is trained to accurately predict beam sizes based on insertion gaps, the comparisons on vertical beam sizes (top) and deviations of different model predictions (bottom) clearly shows that the neural network can outperform regression models Subfigures **c.** and **d.** are reproduced from Leeman *et al.*[190].

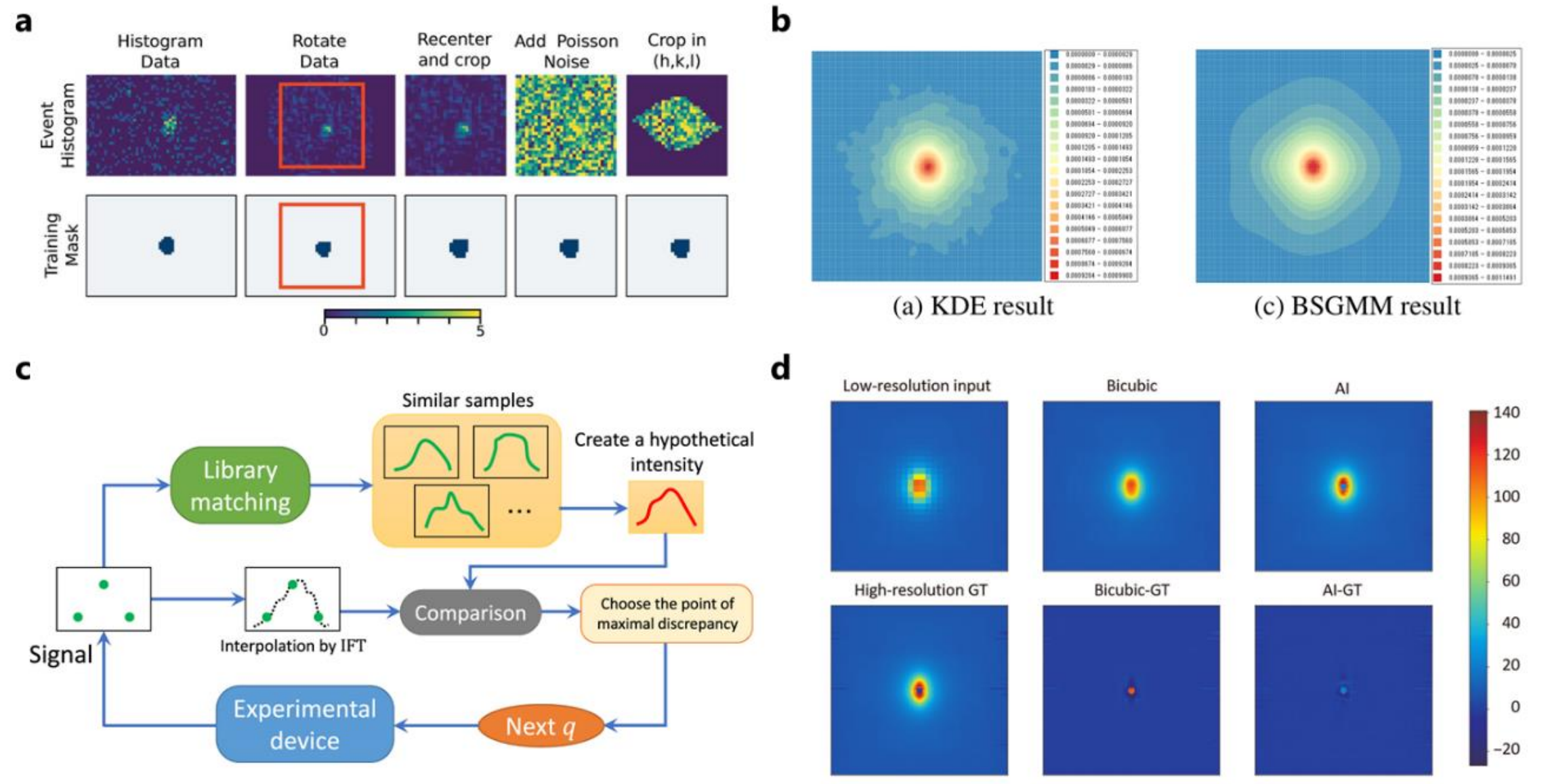

**Figure 12. Machine learning applications in collecting and processing scattering data. a.** Deep U-Net is applied to provide better peak masks for more accurate Bragg peaks integral. Reproduced from Sullivan *et al*.[84]. **b.** Predicted SANS pattern from traditional kernel density estimation (KDE) and Gaussian mixed model with B-spline-based prior (BSGMM), where the latter approach yields smoother, suggesting better predictions are obtained. Reproduced from Asahara *et al*.[193]. **c.** Data-driven sequential measurement for SANS by proposing next sampling **Q** points based on previously measured data. Reproduced from Kanazawa *et al.*[194]. **d.** Comparisons about super-resolution for SANS data made by CNN-based method and baseline bicubic up-sampling algorithm, where the CNN-based method yields better reconstruction of the high-resolution data. Reproduced from Chang *et al.*[107].